*Review*

# Position and Speed Control of Brushless DC Motors Using Sensorless Techniques and Application Trends

**José Carlos Gamazo-Real *, Ernesto Vázquez-Sánchez and Jaime Gómez-Gil**

Department of Signal Theory, Communications and Telematic Engineering, University of Valladolid (UVA), 47011 Valladolid, Spain; E-Mails: ernesto.vazquez@uva.es (E.V.-S.); jgomez@tel.uva.es (J.G.-G.)

* Author to whom correspondence should be addressed; E-Mail: jcgamazo@ribera.tel.uva.es; Tel.: +34-983-185-556; Fax: +34-983-423-667.

**Abstract:** This paper provides a technical review of position and speed sensorless methods for controlling Brushless Direct Current (BLDC) motor drives, including the background analysis using sensors, limitations and advances. The performance and reliability of BLDC motor drivers have been improved because the conventional control and sensing techniques have been improved through sensorless technology. Then, in this paper sensorless advances are reviewed and recent developments in this area are introduced with their inherent advantages and drawbacks, including the analysis of practical implementation issues and applications. The study includes a deep overview of state-of-the-art back-EMF sensing methods, which includes Terminal Voltage Sensing, Third Harmonic Voltage Integration, Terminal Current Sensing, Back-EMF Integration and PWM strategies. Also, the most relevant techniques based on estimation and models are briefly analysed, such as Sliding-mode Observer, Extended Kalman Filter, Model Reference Adaptive System, Adaptive observers (Full-order and Pseudoreduced-order) and Artificial Neural Networks.

## 1. Introduction

For the past two decades several Asian countries such as Japan, which have been under pressure from high energy prices, have implemented variable speed PM motor drives for energy saving applications such as air conditioners and refrigerators [1]. On the other hand, the U.S.A. has kept on using cheap induction motor drives, which have around 10% lower efficiency than adjustable PM motor drives for energy saving applications. Therefore recently, the increase in energy prices spurs higher demands of variable speed PM motor drives. Also, recent rapid proliferation of motor drives into the automobile industry, based on hybrid drives, generates a serious demand for high efficient PM motor drives, and this was the beginning of interest in BLDC motors.

BLDC motors, also called Permanent Magnet DC Synchronous motors, are one of the motor types that have more rapidly gained popularity, mainly because of their better characteristics and performance [2]. These motors are used in a great amount of industrial sectors because their architecture is suitable for any safety critical applications.

The brushless DC motor is a synchronous electric motor that, from a modelling perspective, looks exactly like a DC motor, having a linear relationship between current and torque, voltage and rpm. It is an electronically controlled commutation system, instead of having a mechanical commutation, which is typical of brushed motors. Additionally, the electromagnets do not move, the permanent magnets rotate and the armature remains static. This gets around the problem of how to transfer current to a moving armature. In order to do this, the brush-system/commutator assembly is replaced by an intelligent electronic controller, which performs the same power distribution as a brushed DC motor [3]. BLDC motors have many advantages over brushed DC motors and induction motors, such as a better speed *versus* torque characteristics, high dynamic response, high efficiency and reliability, long operating life (no brush erosion), noiseless operation, higher speed ranges, and reduction of electromagnetic interference (EMI). In addition, the ratio of delivered torque to the size of the motor is higher, making it useful in applications where space and weight are critical factors, especially in aerospace applications.

The control of BLDC motors can be done in sensor or sensorless mode, but to reduce overall cost of actuating devices, sensorless control techniques are normally used. The advantage of sensorless BLDC motor control is that the sensing part can be omitted, and thus overall costs can be considerably reduced. The disadvantages of sensorless control are higher requirements for control algorithms and more complicated electronics [3]. All of the electrical motors that do not require an electrical connection (made with brushes) between stationary and rotating parts can be considered as brushless permanent magnet (PM) machines [4], which can be categorised based on the PMs mounting and the back-EMF shape. The PMs can be *surface mounted on the rotor* (SMPM) or installed *inside of the rotor* (IPM) [5], and the back-EMF shape can either be sinusoidal or trapezoidal. According to the back-EMF shape, *PM AC synchronous motors* (PMAC or PMSM) have sinusoidal back-EMF and *Brushless DC motors* (BLDC or BPM) have trapezoidal back-EMF. A PMAC motor is typically excited by a three-phase sinusoidal current, and a BLDC motor is usually powered by a set of currents having a quasi-square waveform [6,7].

Because of their high power density, reliability, efficiency, maintenance free nature and silent operation, permanent magnet (PM) motors have been widely used in a variety of applications in industrial automation, computers, aerospace, military (gun turrets drives for combat vehicles) [3], automotive (hybrid vehicles) [8] and household products. However, the PM BLDC motors are inherently electronically controlled and require rotor position information for proper commutation of currents in its stator windings. It is not desirable to use the *position sensors* for applications where reliability is of utmost importance because a sensor failure may cause instability in the control system. These limitations of using position sensors combined with the availability of powerful and economical microprocessors have spurred the development of sensorless control technology. Solving this problem effectively will open the way for full penetration of this motor drive into all low cost, high reliability, and large volume applications.

The remainder of the paper is arranged as follows. Section 2 describes the position and speed control fundamentals of BLDC motors using sensors. Next, Section 3 explains the control improvements applying sensorless techniques, describing the motor controller model and the most important techniques based on back-EMF sensing. Section 4 also briefly analyses the sensorless techniques using estimators and model-based schemes. In addition, Section 5 compares the feasibility of the control methods, and describes some relevant implementation issues, such as open-loop starting. Finally, Section 6 provides an overview for the applications of BLDC motor controllers, as well as conclusions are drawn in Section 7.

## 2. Position and Speed Control of BLDC Motors Using Sensors

PM motor drives require a rotor position sensor to properly perform phase commutation and/or current control. For PMAC motors, a constant supply of position information is necessary; thus a position sensor with high resolution, such as a *shaft encoder or a resolver*, is typically used. For BLDC motors, only the knowledge of six phase-commutation instants per electrical cycle is needed; therefore, low-cost *Hall-effect sensors* are usually used. Also, *electromagnetic variable reluctance* (VR) *sensors* or *accelerometers* have been extensively applied to measure motor position and speed. The reality is that angular motion sensors based on magnetic field sensing principles stand out because of their many inherent advantages and sensing benefits.

### *2.1. Position and Speed Sensors*

As explained before, some of the most frequently used devices in position and speed applications are Hall-effect sensors, variable reluctance sensors and accelerometers. Each of these types of devices is discussed further below.

#### 2.1.1. Hall-effect sensors

These kinds of devices are based on Hall-effect theory, which states that if an electric current-carrying conductor is kept in a magnetic field, the magnetic field exerts a transverse force on the moving charge carriers that tends to push them to one side of the conductor. A build-up of charge at the sides of the conductors will balance this magnetic influence producing a measurable voltage between the two sides of the conductor. The presence of this measurable transverse voltage is called the Hall-effect because it was discovered by Edwin Hall in 1879.

Unlike a brushed DC motor, the commutation of a BLDC motor is controlled electronically. To rotate the BLDC motor the stator windings should be energized in a sequence. It is important to know the rotor position in order to understand which winding will be energized following the energizing sequence. Rotor position is sensed using Hall-effect sensors embedded into the stator [9].

Most BLDC motors have three Hall sensors inside the stator on the non-driving end of the motor. Whenever the rotor magnetic poles pass near the Hall sensors they give a high or low signal indicating the N or S pole is passing near the sensors. Based on the combination of these three Hall sensor signals, the exact sequence of commutation can be determined. Figure 1 shows a transverse section of a BLDC motor with a rotor that has alternate N and S permanent magnets. Hall sensors are embedded into the stationary part of the motor. Embedding the Hall sensors into the stator is a complex process because any misalignment in these Hall sensors with respect to the rotor magnets will generate an error in determination of the rotor position. To simplify the process of mounting the Hall sensors onto the stator some motors may have the Hall sensor magnets on the rotor, in addition to the main rotor magnets. Therefore, whenever the rotor rotates the Hall sensor magnets give the same effect as the main magnets. The Hall sensors are normally mounted on a printed circuit board and fixed to the enclosure cap on the non-driving end. This enables users to adjust the complete assembly of Hall sensors to align with the rotor magnets in order to achieve the best performance [10].

**Figure 1.** BLDC motor transverse section [10].

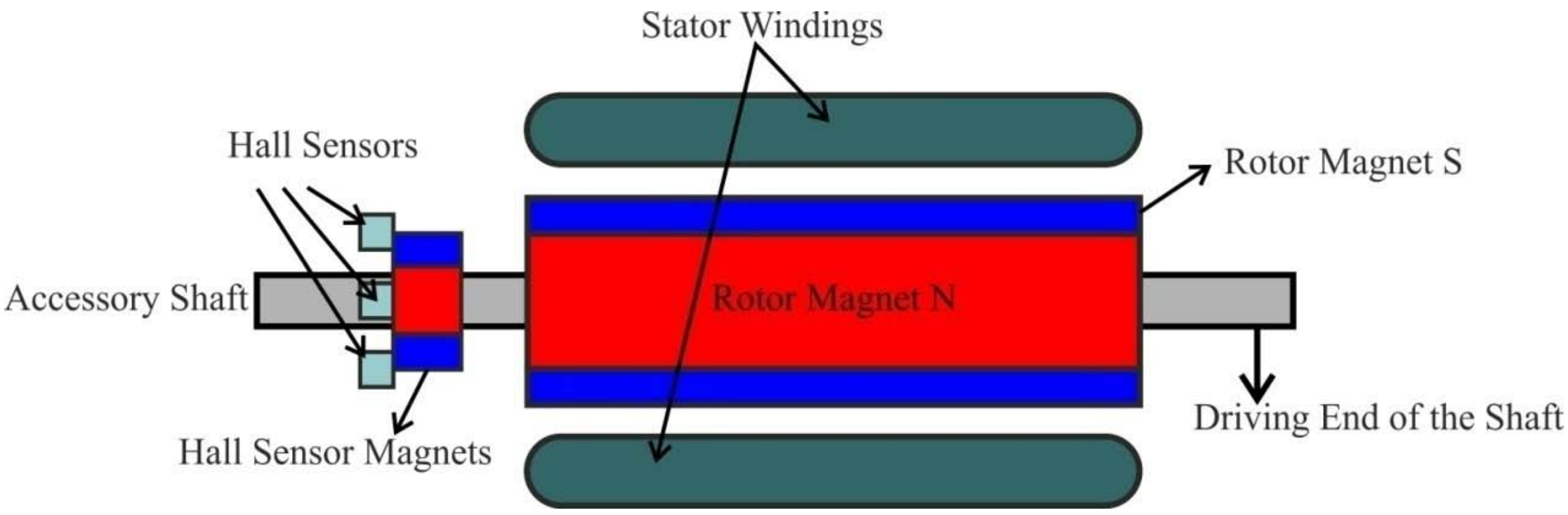

Nowadays, because miniaturized brushless motors are introduced in many applications, new position sensors are being developed, such as a three branches vertical Hall sensor [11] depicted in Figure 2a. The connecting principle between the brushless motor and this sensor is reminiscent of the miniaturized magnetic angular encoder based on 3-D Hall sensors. A permanent magnet is fixed at the end of a rotary shaft and the magnetic sensor is placed below, and the magnet creates a magnetic field parallel to the sensor surface. This surface corresponds to the sensitive directions of the magnetic sensor. Three-phase brushless motors need three signals with a phase shift of 120° for control, so a closed-loop regulation may be used to improve the motor performance. Each branch could be interpreted as a half of a vertical Hall sensor and are rotated by 120° in comparison to the other. Only a half of a vertical Hall sensor is used since little space is available for the five electrical contacts (two for the supply voltage and three to extract the Hall voltages). This sensor automatically creates three signals with a phase shift of 120°, which directly correspond to the motor driving signals, to simplify the motor control in a closed-loop configuration. A drawing of this device's use as angular position sensor for brushless motor control is given in Figure 2b. A first alignment is between the rotor orientation and the permanent magnet, and a second alignment is between the stator and the sensor. This alignment will allow the motion information for the motor and the information about its shaft angular position.

**Figure 2.** **(a)** Schematic representation of a three branches Hall sensor. **(b)** Three branches vertical Hall device mounted as angular position sensor for brushless motor control [11].

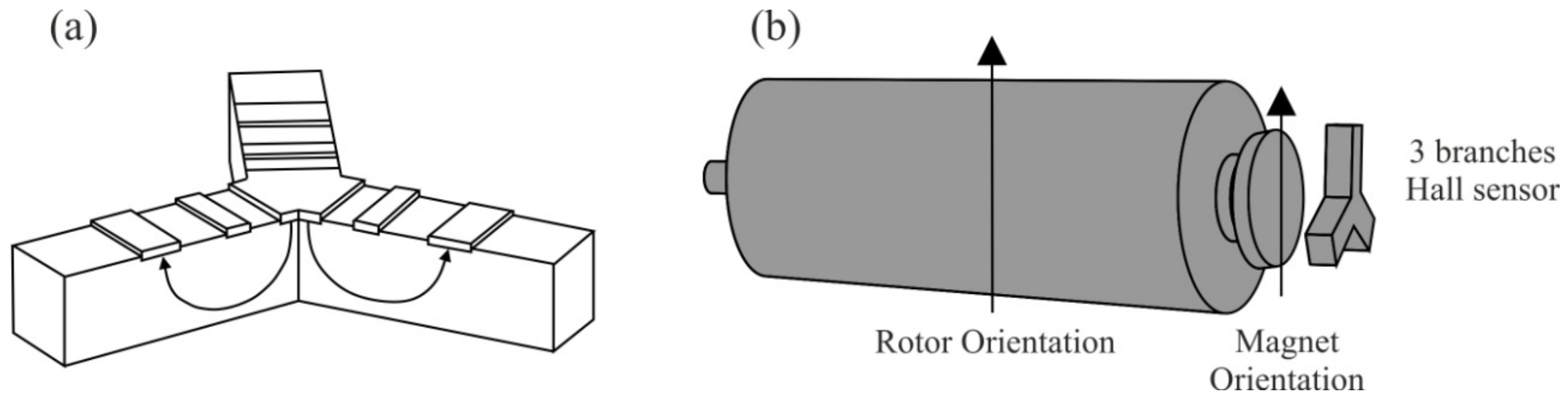

2.1.2. Variable reluctance (VR) wheel speed sensors

This kind of sensor is used to measure position and speed of moving metal components, and is often referred as a passive magnetic sensor because it does not need to be powered. It consist of a permanent magnet, a ferromagnetic pole piece, a pickup coil, and a rotating toothed wheel, as Figure 3 illustrates. This device is basically a permanent magnet with wire wrapped around it. It is usually a simple circuit of only two wires where in most cases polarity is not important, and the physics behind its operation include magnetic induction [12].

**Figure 3.** Variable Reluctance sensor that senses movement of the toothed wheel [12].

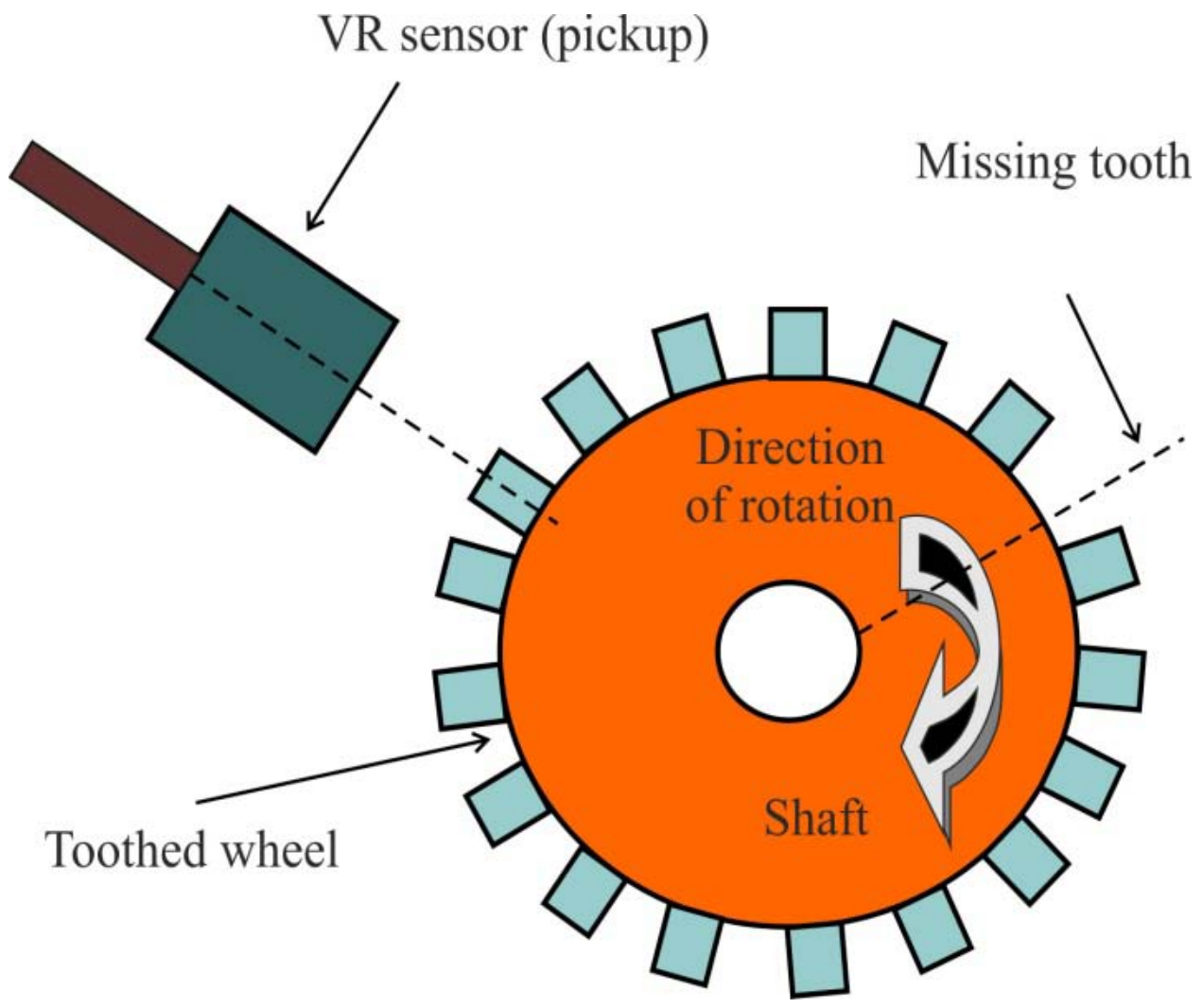

As the teeth pass through the sensor's magnetic field, the amount of magnetic flux passing through the permanent magnet varies. When the tooth gear is close to the sensor, the flux is at maximum. When the tooth is further away, the flux drops off. The moving target results in a time-varying flux that induces a voltage in the coil, producing an electrical analog wave. *The frequency and voltage of the* *analog signal is proportional to velocity of the rotating toothed wheel.* Each passing discontinuity in the target causes the VR sensor to generate a pulse. The cyclical pulse train or a digital waveform created can be interpreted by the BLDC motor controller.

The advantages of the VR sensor can be summarized as follows: low cost, robust proven speed and position sensing technology (it can operate at temperatures in excess of 300 °C), self-generating electrical signal which requires no external power supply, fewer wiring connections which contribute to excellent reliability, and a wide range of output, resistance, and inductance requirements so that the device can be tailored to meet specific control requirements [12].

Due to the fact that these sensors are very small, they can be embedded in places where other sensors may not fit. For instance, when sealed in protective cases they can be resistant to high temperatures and high pressures, as well as chemical attacks [13]. Through the monitoring of the health of running motors, severe and unexpected motor failures can be avoided and control system reliability and maintainability can be improved. If the VR was integrated inside a motor case for an application in a harsh environment, sensor cables could be easily damaged in that environment. Then, a wireless and powerless sensing solution should be applied using electromagnetic pulses for passing through the motor casing to deliver the sensor signal to the motor controller [14].

The *Hall-effect sensor* explained before is an alternative but more expensive technology, so *VR sensors* are the most suitable choice to measure the rotor position and speed.

2.1.3. Accelerometers

An accelerometer is a electromechanical device that measures acceleration forces, which are related to the freefall effect. Several types are available to detect magnitude and direction of the acceleration as a vector quantity, and can be used to sense position, vibration and shock. The most common design is based on a combination of Newton's law of mass acceleration and Hooke's law of spring action [13]. Then, conceptually, an accelerometer behaves as a damped mass on a spring, which is depicted in Figure 4. When the accelerometer experiences acceleration, the mass is displaced to the point that the spring is able to accelerate the mass at the same rate as the casing. The displacement is then measured to give the acceleration.

**Figure 4.** Basic spring-mass system accelerometer [13].

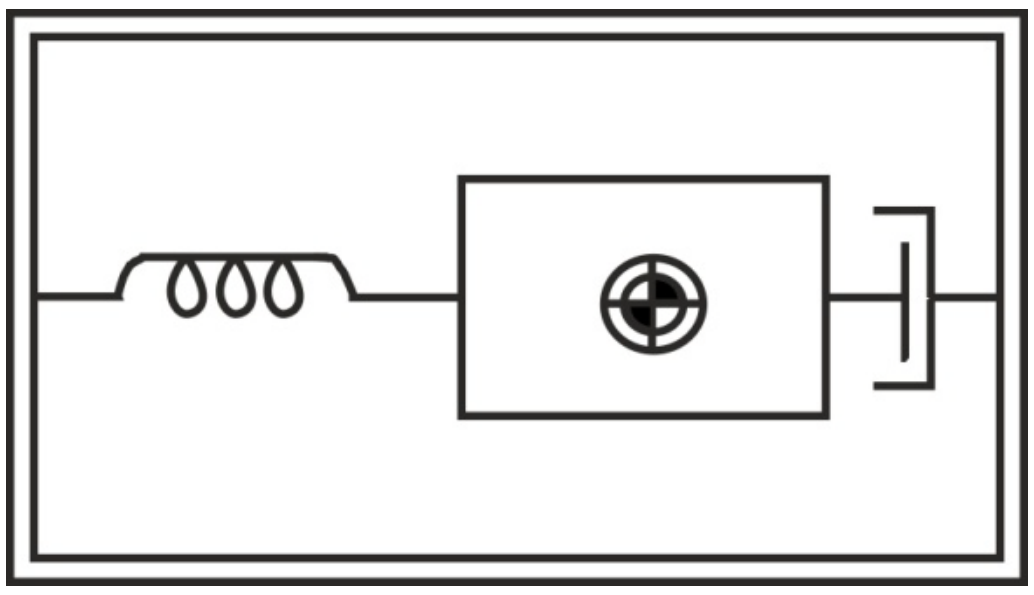

Under steady-state conditions, the measurement of acceleration is reduced to a measurement of spring extension (linear displacement) showed in Equation (1):

$$a = \frac{k}{m} \cdot \Delta x \quad (1)$$

where *k* is the spring constant, *m* is the seismic (or proof) mass, and *Δx* is the distance that is stretched the spring from its equilibrium position with a force given by Equation (2), which is described by Newton's and Hooke's laws:

$$F = k \cdot \Delta x \quad (2)$$

There is a wide variety of accelerometers depending on the requirements of natural frequency, damping, temperature, size, weight, hysteresis, and so on. Some of these types are piezoelectric, piezoresistive, variable capacitance, linear variable differential transformers (LVDT), potentiometric, among many others [13].

Modern accelerometers are often small micro electro-mechanical systems (MEMS), and are indeed the simplest MEMS devices possible, and consist of little more than a cantilever beam with a proof mass. The MEMS accelerometer is silicon micro-machined, and therefore, can be easily integrated with the signal processing circuits [14]. This is different when compare with a traditional accelerometer such as the piezoelectric kind.

### 2.2. Conventional Control Method Using Sensors

A BLDC motor is driven by voltage strokes coupled with the rotor position. These strokes must be properly applied to the active phases of the three-phase winding system so that the angle between the stator flux and the rotor flux is kept close to 90° to get the maximum generated torque. Therefore, the controller needs some means of determining the rotor's orientation/position (relative to the stator coils), such as Hall-effect sensors, which are mounted in or near the machine's air gap to detect the magnetic field of the passing rotor magnets. Each sensor outputs a high level for 180° of an electrical rotation, and a low level for the other 180°. The three sensors have a 60° relative offset from each other. This divides a rotation into six phases (3-bit code) [9].

**Figure 5.** Electronically commutated BLDC motor drive [16].

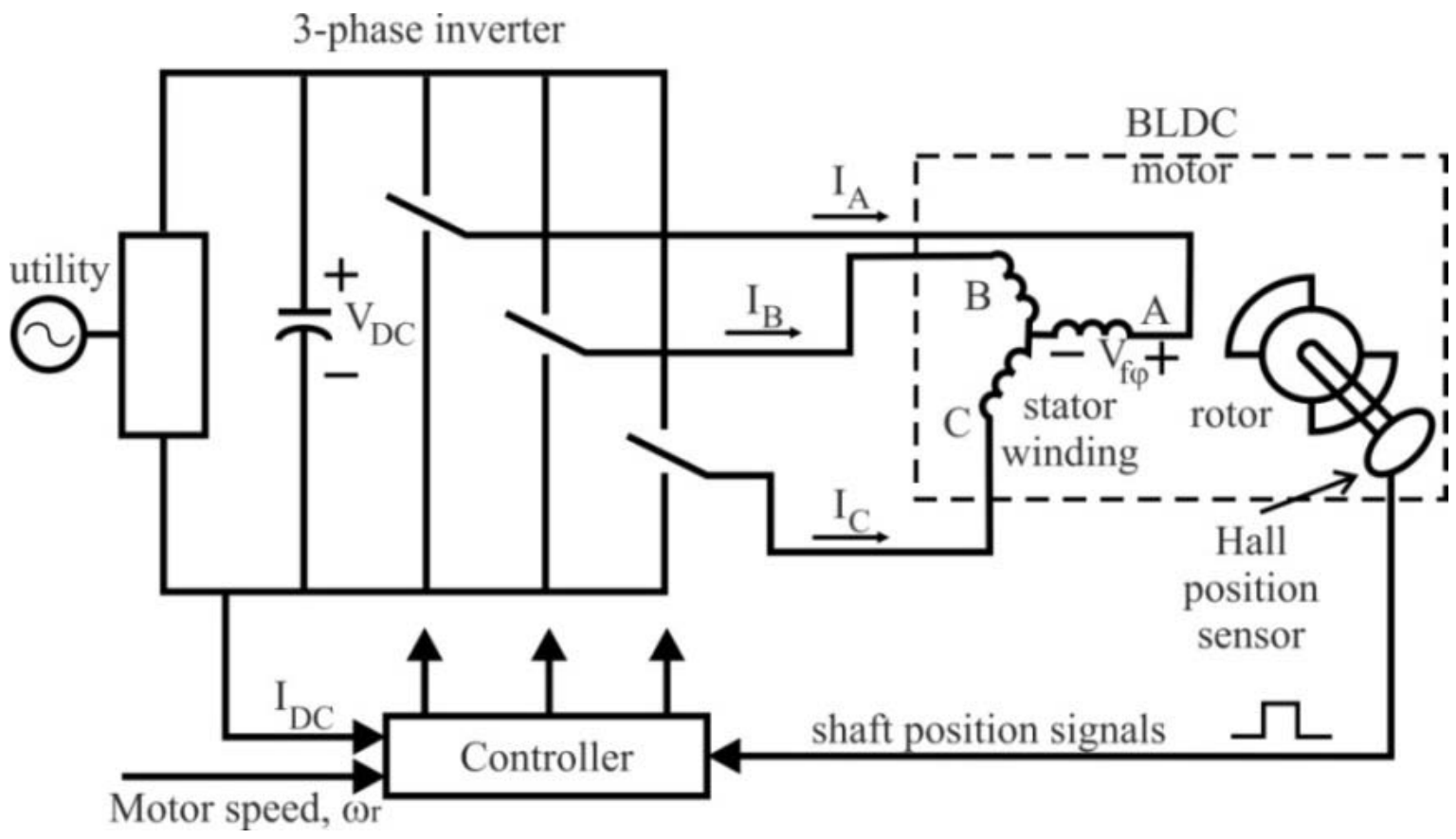

The process of switching the current to flow through only two phases for every 60 electrical degree rotation of the rotor is called *electronic commutation*. The motor is supplied from a three-phase inverter, and the switching actions can be simply triggered by the use of signals from position sensors that are mounted at appropriate points around the stator. When mounted at 60 electrical degree intervals and aligned properly with the stator phase windings these Hall switches deliver digital pulses that can be decoded into the desired three-phase switching sequence [15]. A BLDC motor drive with a six-step inverter and Hall position sensors is shown in Figure 5.

**Figure 6.** Hall sensor signal, back-EMF, output torque and phase current [10].

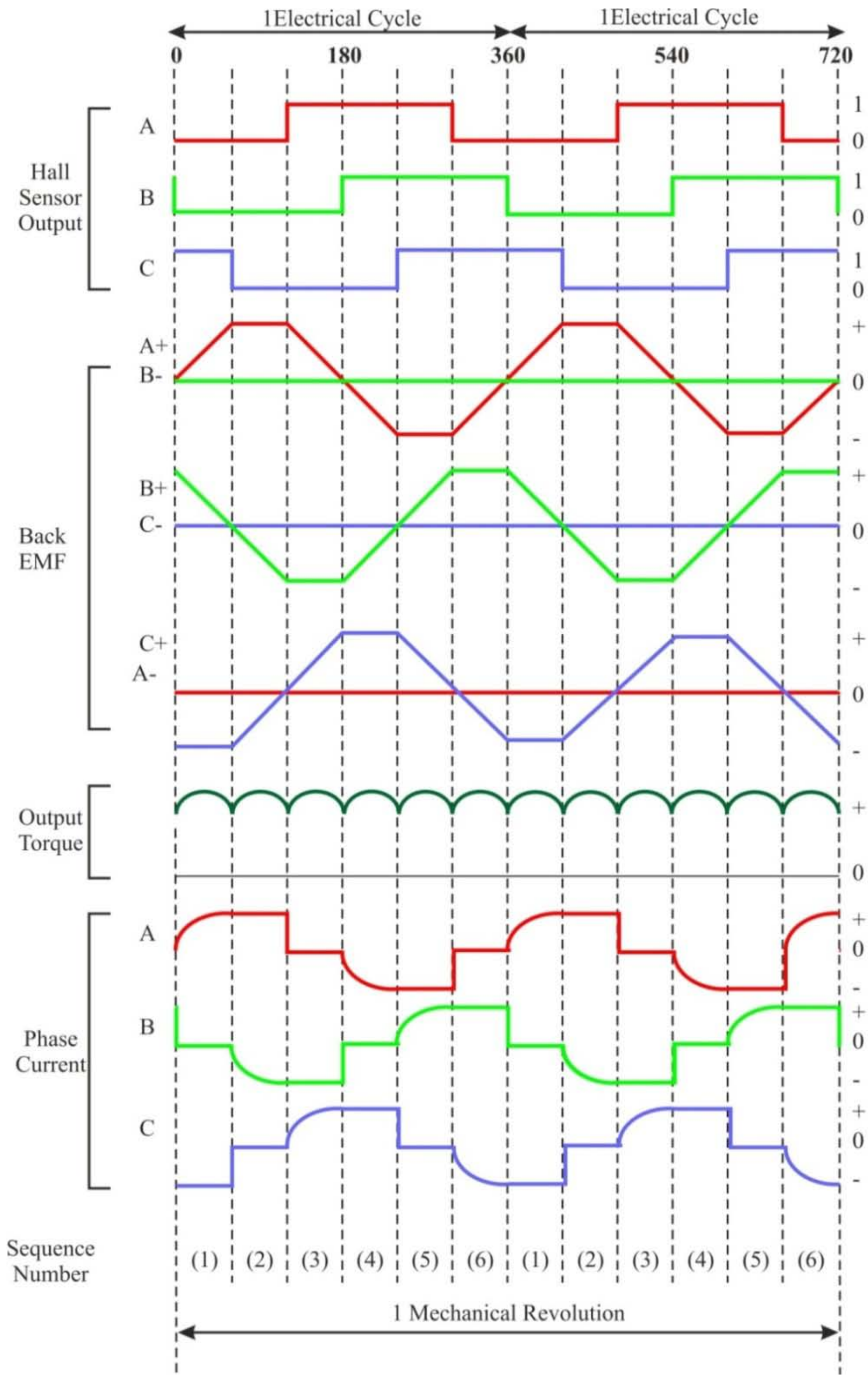

Such a drive usually also has a current loop to regulate the stator current, and an outer speed loop for speed control [16]. The speed of the motor can be controlled if the voltage across the motor is changed, which can be achieved easily varying the duty cycle of the PWM signal used to control the six switches of the three-phase bridge.

Only two inverter switches, one in the upper inverter bank and one in the lower inverter bank, are conducting at any instant. These discrete switching events ensure that the sequence of conducting pairs of stator terminals is maintained [16]. Figure 6 shows an example of Hall sensor signals with respect to back-EMF and the phase current. One of the Hall sensors changes the state every 60 electrical degrees of rotation. Given this, it takes six steps to complete an electrical cycle. However, one electrical cycle may not correspond to a complete mechanical revolution of the rotor. The number of electrical cycles to be repeated to complete a mechanical rotation is determined by the rotor pole pairs. For each rotor pole pair, one electrical cycle is completed. *The number of electrical cycles/rotations equals the rotor pole pairs* [10]. This sequence of conducting pairs is essential to the production of a constant output torque.

In summary, permanent magnet motor drives require a rotor position sensor to properly perform phase commutation, but there are several drawbacks when such types of position sensors are used. The main drawbacks are the increased cost and size of the motor, and a special arrangement needs to be made for mounting the sensors. Further, Hall sensors are temperature sensitive and hence the operation of the motor is limited, which could reduce the system reliability because of the extra components and wiring [7]. To reduce cost and improve reliability such position sensors may be eliminated. To this end, many sensorless schemes have been reported for position (and speed) control of BLDC motors [6].

## 3. Techniques and Advances in Sensorless Control

Position sensors can be completely eliminated, thus reducing further cost and size of motor assembly, in those applications in which only variable speed control (*i.e.*, no positioning) is required and system dynamics is not particularly demanding (*i.e.*, slowly or, at least, predictably varying load). In fact, some control methods, such as back-EMF and current sensing, provide, in most cases, enough information to estimate with sufficient precision the rotor position and, therefore, to operate the motor with synchronous phase currents. A PM brushless drive that does not require position sensors but only electrical measurements is called a *sensorless drive* [4].

The BLDC motor provides an attractive candidate for sensorless operation because the nature of its excitation inherently offers a low-cost way to extract rotor position information from motor-terminal voltages. In the excitation of a three-phase BLDC motor, except for the phase-commutation periods, only two of the three phase windings are conducting at a time and the no conducting phase carries the back-EMF. There are many categories of sensorless control strategies [6]; however, the most popular category is based on back electromotive forces or back-EMFs [17]. Sensing back-EMF of unused phase is the most cost efficient method to obtain the commutation sequence in star wound motors. Since back-EMF is zero at standstill and proportional to speed, the measured terminal voltage that has large signal-to-noise ratio cannot detect zero crossing at low speeds. That is the reason why in all back-EMF-based sensorless methods the low-speed performance is limited, and an *open-loop starting strategy* is required [18].

Generally, a brushless DC motor consists of a permanent magnet synchronous motor that converts electrical energy to mechanical energy, an inverter corresponding to brushes and commutators, and a *shaft position sensor* [19], as Figure 7 shows. In this figure, each of the three inverter phases are highlighted in a different colour, including the neutral point: red phase *A*, green phase *B*, blue phase *C*, and pink neutral point *N*. The stator iron of the BLDC motor has a non-linear magnetic saturation characteristic, which is the basis from which it is possible to determine the initial position of the rotor. When the stator winding is energized, applying a DC voltage for a certain time, a magnetic field with a fixed direction will be established. Then, the current responses are different due to the inductance difference, and this variation of the current responses contains the information of the rotor position [20]. Therefore, *the inductance of stator winding is a function of the rotor position*.

**Figure 7.** Typical sensorless BLDC motor drive [19].

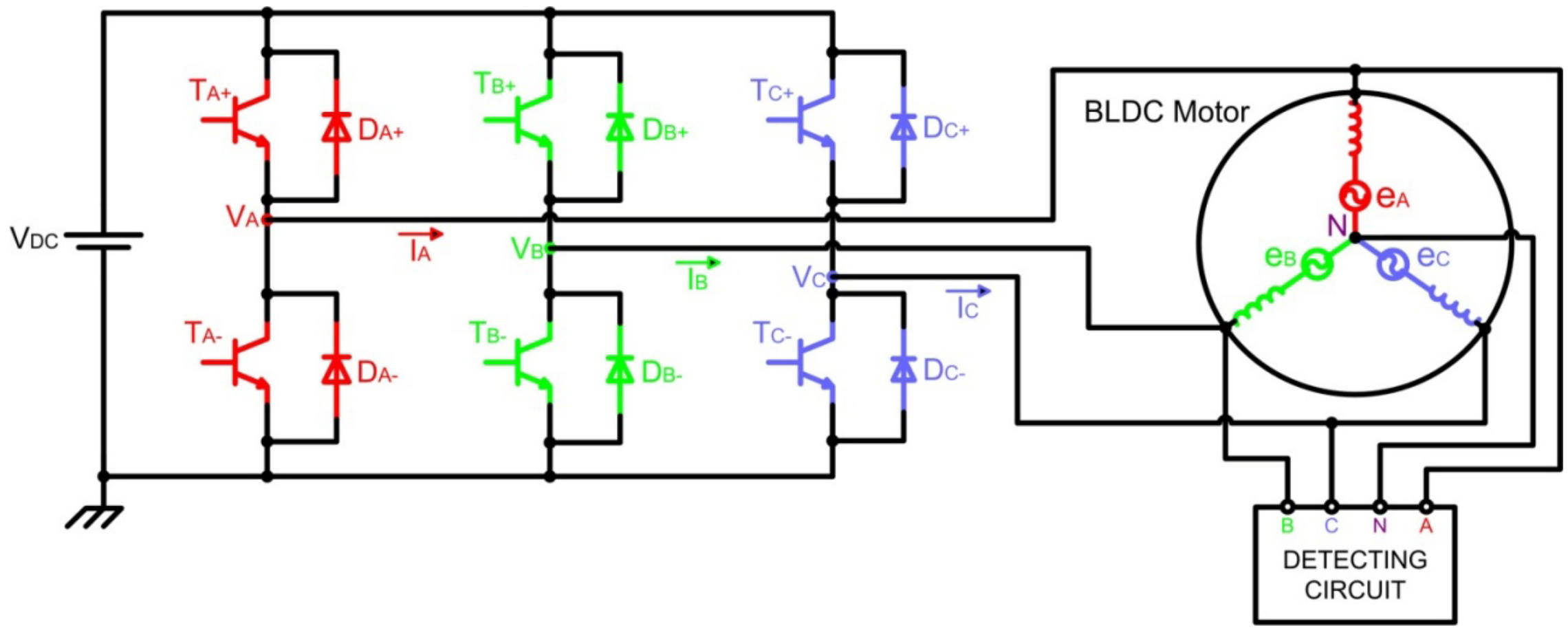

The analysis of the circuit depicted in Figure 7 is based on the motor model for phase *A* (highlighted in red colour), illustrated in Figure 8, and the following assumptions are considered [21]:

- The motor is not saturated.
- Stator resistances of all the windings are equal ($R_S$), self inductances are constant ($L_S$) and mutual inductances ($M$) are zero.
- Iron losses are negligible.

**Figure 8**. Equivalent circuit of the BLDC motor for phase *A* [21].

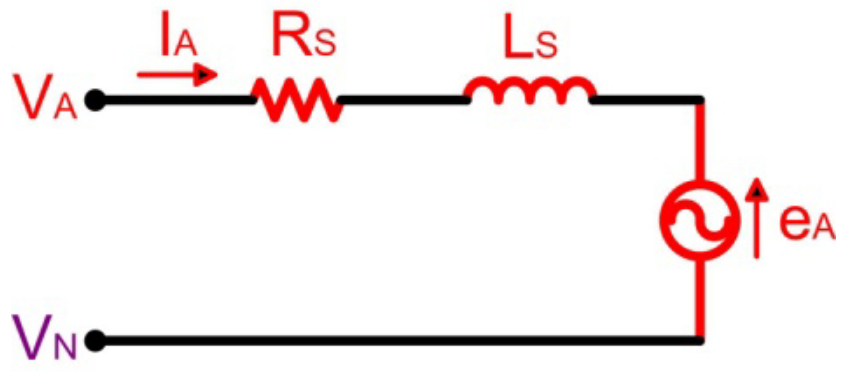

Then, the voltage function of the conducting phase winding might be expressed as indicated in Equation (3):

$$V_{DC} = I \cdot R_S + L_S \cdot \frac{dI}{dt} + e \qquad (3)$$

where $V_{DC}$ is the DC-link voltage, $R_S$ and $L_S$ are the equivalent resistance and inductance of stator phase winding respectively, and $e$ is the trapezoidal shaped back-EMF.

In this paper, conventional and recent advancement of back-EMF sensing methods for the PM BLDC motors and generators are presented, which are split in two categories; direct and indirect back-EMF detection [22]:

- *Direct back-EMF detection methods*: the back-EMF of floating phase is sensed and its zero crossing is detected by comparing it with neutral point voltage. This scheme suffers from high common mode voltage and high frequency noise due to the PWM drive, so it requires low pass filters, and voltage dividers. The methods can be classified as:
  - Back-EMF Zero Crossing Detection (ZCD) or Terminal Voltage Sensing.
  - PWM strategies.
- *Indirect back-EMF detection methods*: because filtering introduces commutation delay at high speeds and attenuation causes reduction in signal sensitivity at low speeds, the speed range is narrowed in direct back-EMF detection methods. In order to reduce switching noise, the indirect back-EMF detection methods are used. These methods are the following:
  - Back-EMF Integration.
  - Third Harmonic Voltage Integration.
  - Free-wheeling Diode Conduction or Terminal Current Sensing.

### *3.1. Back-EMF Zero Crossing Detection method (Terminal Voltage Sensing)*

The zero-crossing approach is one of the simplest methods of back-EMF sensing technique, and is based on detecting the instant at which the back-EMF in the unexcited phase crosses zero. This zero crossing triggers a timer, which may be as simple as an RC time constant, so that the next sequential inverter commutation occurs at the end to this timing interval [23].

For typical operation of a BLDC motor, the phase current and back-EMF should be aligned to generate constant torque. The current commutation point shown in Figure 9 can be estimated by the zero crossing point (ZCP) of back-EMFs and a 30° phase shift [1,4], using a six-step commutation scheme through a three-phase inverter for driving the BLDC motor. The conducting interval for each phase is 120 electrical degrees. Therefore, only two phases conduct current at any time, leaving the third phase floating. In order to produce maximum torque, the inverter should be commutated every 60° by detecting zero crossing of back-EMF on the floating coil of the motor [24], so that current is in phase with the back-EMF.

This technique of delaying 30° (electrical degrees) from zero crossing instant of the back-EMF is not affected much by speed changes [7]. To detect the ZCPs, the phase back-EMF should be monitored during the silent phase (when the particular phase current is zero) and the terminal voltages should be low-pass filtered first.

**Figure 9.** Zero crossing points of the back-EMF and phase current commutation points [25].

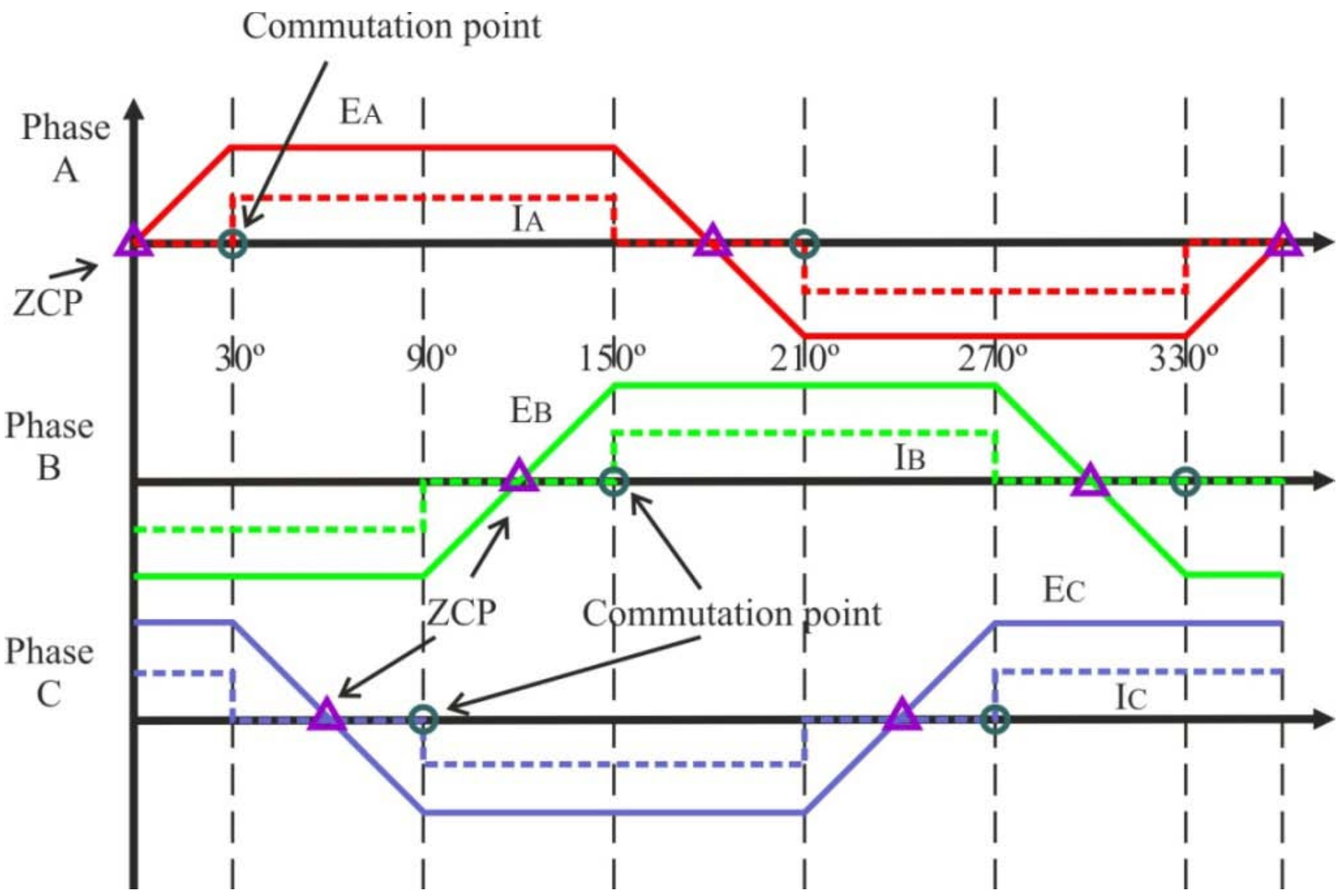

Three low-pass filters (LPFs) are utilized to eliminate higher harmonics in the phase terminal voltages caused by the inverter switching. The time delay of LPFs will limit the high speed operation capability of the BLDC machine [1,24,26]. It's necessary to point out the importance of filters when a BLDC motor drive is designed, which are used to eliminate high frequency components of the terminal voltages and to extract back-EMF of the motor.

The terminal voltage of the opened or floating phase is given by Equation (4):

$$V_C = e_C + V_N = e_C + \frac{V_{CE} - V_F}{2} - \frac{e_A + e_B}{2} \quad (4)$$

where $e_C$ is the back-EMF of the opened phase ($C$), $V_N$ is the potential of the motor neutral point, and $V_{CE}$ and $V_F$ are the forward voltage drop of the transistors and diodes, respectively, which implement the motor inverter of Figure 7, respectively.

As the back-EMF of the two conducting phases (A and B) have the same amplitude but opposite sign [19] the terminal voltage of the floating phase results in Equation (5):

$$e_A = -e_B \Rightarrow V_C = e_C + \frac{V_{CE} - V_F}{2} = e_C + \frac{V_B + V_A}{2} \quad (5)$$

where $V_A$=-$V_F$ (forward current of diode $D_{A-}$) and $V_B$=$V_{CE}$ (collector-emitter voltage of transistor $T_{B-}$)

Since the zero-crossing point detection is done at the end of the PWM on-state and only the high-side of the inverter is chopped, and $V_{CE}$ is similar to $T_{A+}$ and $T_{B-}$ transistors, the final detection formula can be represented by Equation (6):

$$V_{CE}^{A+} \approx V_{CE}^{B-} \Rightarrow V_C = e_C + \frac{V_{CE}^{B-} + V_{DC} - V_{CE}^{A+}}{2} \approx e_C + \frac{V_{DC}}{2} \quad (6)$$

Therefore, *the zero-crossing occurs when the voltage of the floating phase reaches one half of the DC rail voltage*. The reason why the end of the PWM on-state is selected as the zero-crossing detection point is that it is noise free to sample at that moment [20].

On the other hand, instead of using analogue LPFs, a unipolar pulse width modulation (PWM) scheme can be used to measure terminal voltages [27,28]. The difference of the ZCD method between on and off state of the PWM signal can also be taken into account [29,30]. Also, the true phase back-EMF signal could be directly obtained from the motor terminal voltage by properly choosing the PWM and sensing strategy (without the motor neutral point voltage information This would provide advantages such as no sensitivity to switching noise, no filtering required, and good motor performance a wide speed range [24,31].

The price for the simplicity of the zero-crossing method tends to be noise sensitivity in detecting the zero crossing, and degraded performance over wide speed ranges unless the timing interval is programmed as a function of rotor speed [23]. Another drawback is that it is not possible to use the noisy terminal voltage to obtain a switching pattern at low speeds since back-EMF is zero at standstill and proportional to rotor speed. Also, the estimated commutation points have position error during the transient period when the speed is accelerated or decelerated rapidly, especially for a system that has low inertia. With this method, rotor position can be detected typically from 20% of the rated speed, then a reduced speed operating range is normally used, typically around 1,000–6,000 rpm [1].

### 3.1.1. Optimizations

As the rotor position information can be extracted by indirectly sensing the back-EMF from only one of the three motor-terminal voltages for a three-phase motor, it is obvious that sensing each terminal voltage can provide two commutation instants. Measuring the time between these two instants, it is possible to interpolate the other four commutation instants (Figure 9 shows that six commutation points are needed), assuming motor speed does not change significantly over consecutive electrical cycles. Depending on the terminal voltage sensing locations, either a low-pass filter or a band-pass filter is used for position information retrieval. The circuit for sensing the other two terminal voltages can therefore be eliminated, leading to a significant reduction in the component count of the sensing circuit. Also, the ZCD method could be improved if a digital filtering procedure is used to identify the true and false ZCPs of phase back-EMF, which are caused by the terminal voltage spikes due to phase commutations [32].

An indirect way of detecting the ZCP of the back-EMF from the three terminal voltages without using the neutral potential is using the difference of the line voltages [33]. Another modification of the technique is to achieve the sensorless commutation by means of a Phase Locked Loop (PLL) and sensing of the phase winding back-EMF voltages [8]. This PLL has a narrow speed range due to the capabilities of the phase detector, and is sensitive to switching noise. In order to simplify the BLDC driver design, it can be built around a sensorless controller chip ML4425 from Fairchild Semiconductor [34,35].

A sensorless Field Oriented Control (FOC) of brushless motors [36], which is known to be more efficient in terms of torque generation compared to back-EMF zero crossing detection methods, is currently under development, and it could be successfully applied to the design of motor pump units for automotive applications [4].

At low speeds or at standstill, the back-EMF detection method can not be applied well because the back-EMF is proportional to the motor speed. In spite of this problem, a starting procedure can be used to start the motor from standstill [20]. In critical applications, such as the intelligent Electro-Mechanical (EMA) and Electro-Hydraulic (EHA) actuators of aviation systems, it is necessary to ensure correct start-up of the DC motor. Electrical commutation in the first running stage is normally realized by classical PWM signal that drives a transistor power stage (see Figure 7), which is *open-loop control* without any position feedback [3].

At high speeds, the long settling time of a parasitic resonant between the motor inductance and the parasitic capacitance of power devices can cause false zero crossing detection of back-EMF. The solution to this problem is to *detect the back-EMF during on time* at high duty cycle [37], so there is enough time for the resonant transient to settle down. Then, during motor start-up and low speed, it is preferred to use the original scheme since there is no signal attenuation; while at high speed, the system can be switched to the improved back-EMF sensing scheme. With the combination of two detection schemes in one system, the motor can run very well over a *wide speed range* [24,38].

3.1.2. Applications

The terminal voltage sensing method is widely used for low cost industrial applications such as fans, pumps and compressor drives where frequent speed variation is not required. Nevertheless, BLDC motors need a *rotor position sensor,* and this reduces the system ruggedness, complicates the motor configuration and its mass production. This sensor can be has been eliminated through this sensing technique. In spite of the back-EMF being zero at standstill, this technique permits the starting of a separately controlled synchronous motor without a sensor, because the PWM signal generated in the control computer chops the motor voltage by the commutation transistors to control the motor speed [1]. An example is a motor pump unit, developed for commercial vehicle applications, in which control strategy can be based on the back-EMF zero-crossing method, and speed control loop is closed by means of the virtual feedback provided by the commutation point prediction [4].

Another important field is the super high speed motors, which are receiving increasing attentions in various applications such as machine tools, because of the advantages of their small size and light weight at the same power level [39].

*3.2. Third Harmonic Voltage Integration method*

This method utilises the third harmonic of the back-EMF to determine the commutation instants of the BLDC motor. It is based on the fact that in a symmetrical three phase Y-connected motor with trapezoidal air gap flux distribution, the summation of the three stator phase voltages results in the elimination of all polyphase , that is fundamental and all the harmonics components like 5th, 7th, *etc.* [40]. The resulting sum is dominated by the third harmonic component that keeps a constant phase displacement with the fundamental air gap voltage for any load and speed.

An appropriate processing of the third harmonic signal allows the estimation of the rotor flux position and a proper inverter current control. In contrast with indirect sensing methods based on the back-EMF signal, the third harmonic requires only a small amount of filtering. As a result, this method is not sensitive to filtering delays, achieving a high performance for a wide speed range. A superior motor starting performance is also achieved because the third harmonic can be detected at low speeds [41].

Referring to Figure 7, the stator voltage of the BLDC motor for phase *A* can be written similarly to Equation (3), where $V_{DC}=V_A$, $I=I_A$, and $e=e_A$. Equivalent expressions can be obtained for the other two stator phases. The harmonic content of the motor air gap or internal voltages $e_A$, $e_B$ and $e_C$ is a function of the rotor magnets and stator winding configurations [40]. For a full pitch magnet and full pitch stator phase winding, the internal voltages can be represented using the Fourier transform, obtaining many voltage harmonic components. If each phase inductance is constant at any rotor position, from the summation of three-terminal to neutral voltages, the third harmonic of the back-EMF can be measured by Equation (7) [25]:

$$V_{SUM} = V_{AN} + V_{BN} + V_{CN} \approx (e_A + e_B + e_C) \approx 3 \cdot E_3 \cdot \sin(3 \cdot \omega_e \cdot t) \quad (7)$$

The summed terminal voltages contain only the third and the multiples of the third harmonic due to the fact that only zero sequence current components can flow through the motor neutral. To obtain switching instants, the filtered voltage signal which provides the third harmonic voltage component is integrated to estimate the rotor flux linkage, as it is shown in Equation (8):

$$\lambda_{r3} = \int V_{SUM} \cdot dt \quad (8)$$

Figure 10 depicts the motor internal voltage corresponding to phase *A*, $e_A$, the third harmonic signal, $V_{SUM}$, obtained from the summation of the stator phase voltages, the rotor flux third harmonic component $\lambda_{r3}$, the rotor flux $\lambda_r$, and the stator phase currents [40]. In order to obtain maximum torque per ampere, the stator current is kept at 90 electrical degrees with respect to the rotor flux. In addition, *the zero crossings of the rotor flux third harmonic component occur at 60 electrical degrees, exactly at every desired current commutation instant.*

**Figure 10.** Back-EMF, third harmonic voltage, rotor flux and rotor flux fundamental components, and motor phase currents [40].

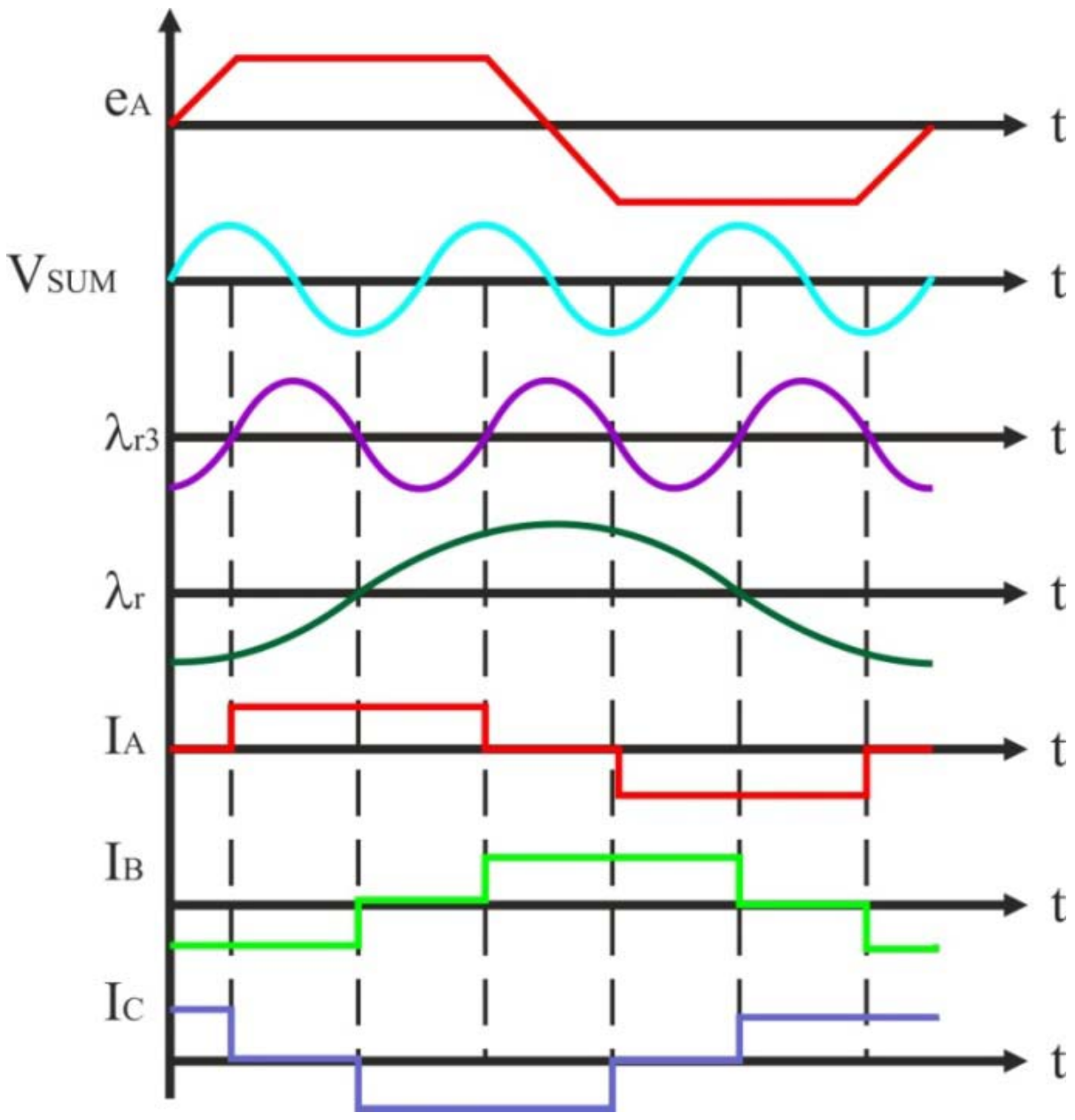

Basically, there are three methods to extract the third harmonic component of the back-EMF, using as reference a permanent-magnet brushless drive with Y-connected resistors to enable the third harmonic component of the back-EMF to be sensed [18]. These methods are the following:

- From the voltage $V_{SN}$ between the star point *S* of the resistor network and the neutral point *N* of the stator windings [42,43].
- From the voltage $V_{SH}$ between *S* and the midpoint *H* of the DC bus [40].
- From the voltage $V_{NH}$ between *N* and *H* [44].

From the foregoing, *only the voltage $V_{SH}$ is suitable for the third harmonic back-EMF sensorless operation of BLDC motors*, but as with all back-EMF based sensorless methods, an open-loop starting procedure has to be employed [18].

#### 3.2.1. Optimizations

The improved version of this method has been developed by using a PLL [17], in which the freewheeling diode conduction period takes place right after the commutation. The inverted terminal voltage measured during the diode conduction period causes position error because of the unbalanced integration, and the commutation angle error was decreased by inverting the measured terminal voltage during the diode conduction period. The method can be integrated into the ML4425 application-specific integrated circuit (ASIC) [25,34,35].

The implementation of this improvement in an experimental system, such as an air compressor, requires low starting torque and the commutation of the BLDC drive is significantly retarded during high-speed operation [17]. To overcome the problem, the ASIC should integrate the third harmonic back-EMF instead of the terminal voltage, such that the commutation retarding is largely reduced and the motor performance is improved [45].

#### 3.2.2. Applications

The key advantages of this technique are simplicity of implementation, low susceptibility to electrical noise, and robustness, what makes it a good option for applications requiring a wide motor speed range. Signal detection at low speeds is possible because the third harmonic signal has a frequency three times higher than the fundamental back-EMF, allowing operation in a wider speed range (100-6,000 rpm) than techniques based on sensing the motor back-EMF [40]. However, at low speed the integration process can cause a serious position error, as noise and offset error from sensing can be accumulated for a relatively long period of time [41].

### *3.3. Free-wheeling Diodes Conduction Detection method (Terminal Current Sensing)*

Up to now, the indirect sensing algorithms explained can be applicable only to the SMPM motors whose winding inductances are almost the same and do not vary with the rotor position. These algorithms, except terminal current sensing method, utilize low pass filters or integration circuits to eliminate PWM frequency noise and to provide a phase delay for correct commutation of the stator current. But, in case of IPM motors, the inductance of stator winding varies with the rotor position. This characteristic introduces unbalance of phase impedances and variation of the potential of the neutral point, and it is impossible to apply the terminal current sensing algorithm. IPM motors are more practical than SMPM motors because of the ruggedness of rotor structure and low inertia [28].

In this technique, the position information can be detected on the basis of the conducting state of free-wheeling diodes connected in antiparallel with power transistors because a current flows in a phase. In this phase any active drive signal is given to the positive and negative side transistors and the current results from the back-EMFs produced in the motor windings. The three-phase permanent magnet synchronous motor has the trapezoidal back-EMFs shown in Figure 9. To produce the maximum torque, the inverter commutation should be performed every 60º so that the rectangular-shaped motor line current is in phase with the back-EMF signal. A starting circuit is needed to give a commutation signal for starting. This approach makes it possible to detect the rotor position over a wide speed range, especially at a lower speed, and to simplify the starting procedure [19].

Therefore, the conducting condition of $D_{C-}$ is given by Equation (9), taking into account that $V_{CE}$ and $V_F$ are much smaller than the back-EMFs. Then, when the back-EMF of phase $C$ ($e_C$) becomes negative, the open-phase current flows through the negative-side diode $D_C$:

$$V_{CE}, V_F << e_A, e_B, e_C \Rightarrow e_C < -\frac{V_{CE} + V_F}{2} \approx 0 \Rightarrow e_C < 0 \quad (9)$$

Since the open-phase current results from the back-EMFs, it is impossible to detect the rotor position at a standstill. Therefore, a suitable starting procedure is necessary to the position sensorless BLDC motor drive. The procedure starts by exciting two arbitrary phases for a preset time. The rotor turns to the direction corresponding to the excited phases. At the end of the preset time, the *open-loop commutation* advancing the switching pattern by 120° is done, and the polarity of the motor line current is altered. After the starting procedure, the motor line current indicates that satisfactory sensorless commutations are performed by the free-wheeling diode conduction method [19].

#### 3.3.1. Applications

This method has a position error of commutation points in the transient state as other back-EMF based methods. But, the most serious drawback of this method is the use of six isolated power supplies for the comparator circuitry to detect current flowing in each freewheeling diode, which prohibits this method from practical applications. However, this technique outperforms the previous back-EMF methods at low-speeds.

### *3.4. Back-EMF Integration Method*

In this technique, the commutation instant is determined by integration of the silent phase's back-EMF (that is the unexcited phase's back-EMF). The main characteristic is that the integrated area of the back-EMFs shown in Figure 11 is approximately the same at all speeds. The integration starts when the silent phase's back-EMF crosses zero. When the integrated value reaches a pre-defined threshold value, which corresponds to a commutation point, the phase current is commutated. If flux weakening operation is required, current advance can be achieved by changing the threshold voltage. The integration approach is less sensitive to switching noise and automatically adjusts for speed changes, but low speed operation is poor due to the error accumulation and offset voltage problems from the integration [25]. As the back-EMF is assumed to vary linearly from positive to negative (trapezoidal back-EMF assumed), and this linear slope is assumed speed-insensitive, the threshold voltage is kept constant throughout the speed range.

**Figure 11.** Integrated areas of the back-EMF [25].

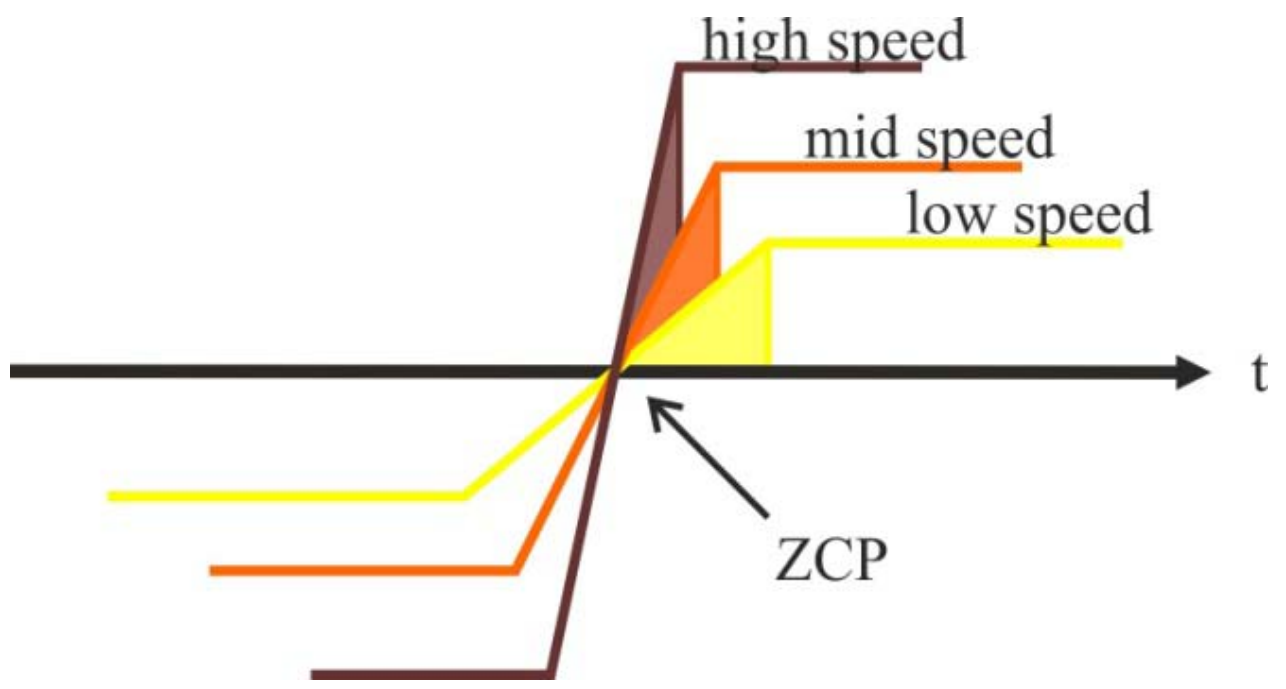

Once the integrated value reaches the threshold voltage, a reset signal is asserted to zero the integrator output. To prevent the integrator from starting to integrate again, the reset signal is kept on long enough to insure that the integrator does not start until the residual current in the open phase has passed a zero-crossing.

The use of discrete *current sensors* for each motor phase will provide complete current feedback, but the cost associated with individual current sensors (e.g., current transformers or Hall-effect sensors) is often prohibitive. An appealing alternative is the use of current sensors which are integrated into the power switches, such as power MOSFET'S and IGBT's, which are available from several device manufacturers with ratings up to several hundreds of volts and several tens of amps. However, embedded current sensors impose their own constraints; for example, the current sensing terminal is not electrically isolated from the associated power device. Also, the availability of new power integrated circuits makes it possible to take more complete advantage of these sensors for the combined purposes of current regulation and overcurrent protection [46].

Finally, the back-EMF integration approach provides significantly improved performance compared to the zero-crossing algorithm explained before. Instead of using the zero-crossing point of the back-EMF waveform to trigger a timer, the rectified back-EMF waveform is fed to an integrator, whose output is compared to pre-set threshold. The adoption of an integrator provides dual advantages of reduced switching noise sensitivity and automatic adjustment of the inverter switching instants according to changes in rotor speed [23].

### *3.5. Methods based on PWM strategies*

There are many methods based on PWM control schemes, but the most relevant are conventional 120º, elimination of virtual neutral point, techniques for low speed, high speed and small power applications, and direct current controlled, which are explained below.

3.5.1. Conventional 120° PWM technique

The block diagram for a three-phase BLDC drive, which consists of a three-phase inverter and a BLDC motor, was shown in Figure 7. It can be controlled by the PWM technique to give proper commutations so that two of the three phases are with on states and the remaining one is with floating state. Moreover, the sequence of commutations is retained in proper order such that the inverter performs the functions of brush and commutator in a conventional DC motor, to generate a rotational stator flux [19,47]. Figure 12 shows the PWM waveforms for this conventional approach [48], which has low switching losses in the inverter side at the cost of significantly high harmonic contents. This results in increase of loss in the motor side [22].

**Figure 12.** PWM waveforms for a conventional approach [48].

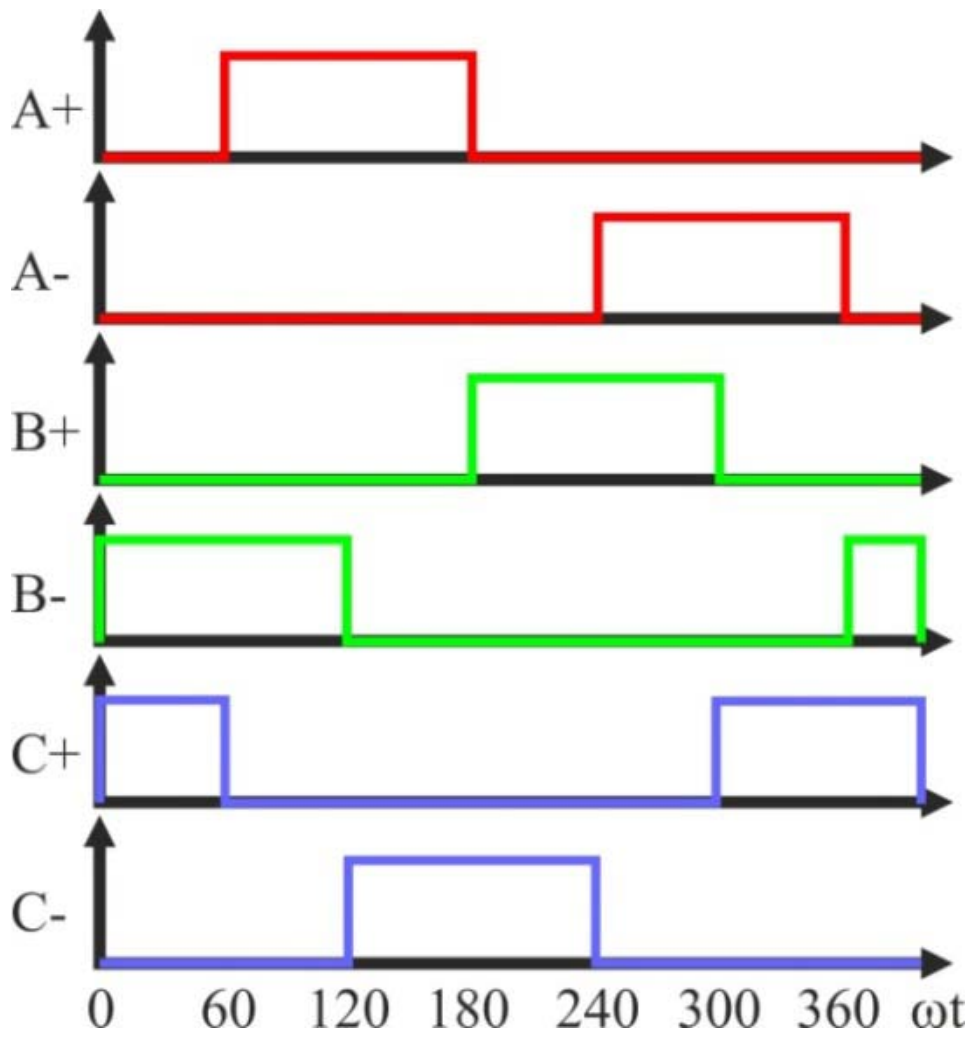

3.5.2. Technique of virtual neutral point elimination

In a typical inverter configuration, as Figure 7 illustrates, two phases are always conducting current and one phase is only available to measure back-EMF. To measure the back-EMF across a phase, the conventional method requires monitoring the phase terminal and the motor neutral point [24], as shown in Figure 13. The zero crossing of the back-EMF can be obtained by comparing the terminal voltage to the neutral point. In most cases, the motor neutral point is not available. The most commonly used method is to build a virtual neutral point that will be theoretically at the same potential as the neutral point of the wye-wound motor [38].

The conventional detection scheme is quite simple and when a PWM signal is used to regulate motor speed or torque/current, the virtual neutral point fluctuates at the PWM frequency. As a result, there is a very high common-mode voltage and high-frequency noise. Voltage dividers and low-pass filters, as shown in Figure 13, are required to reduce the common-mode voltage and minimize the high-frequency noise [38].

**Figure 13.** Back-EMF sensing based on virtual neutral point [38].

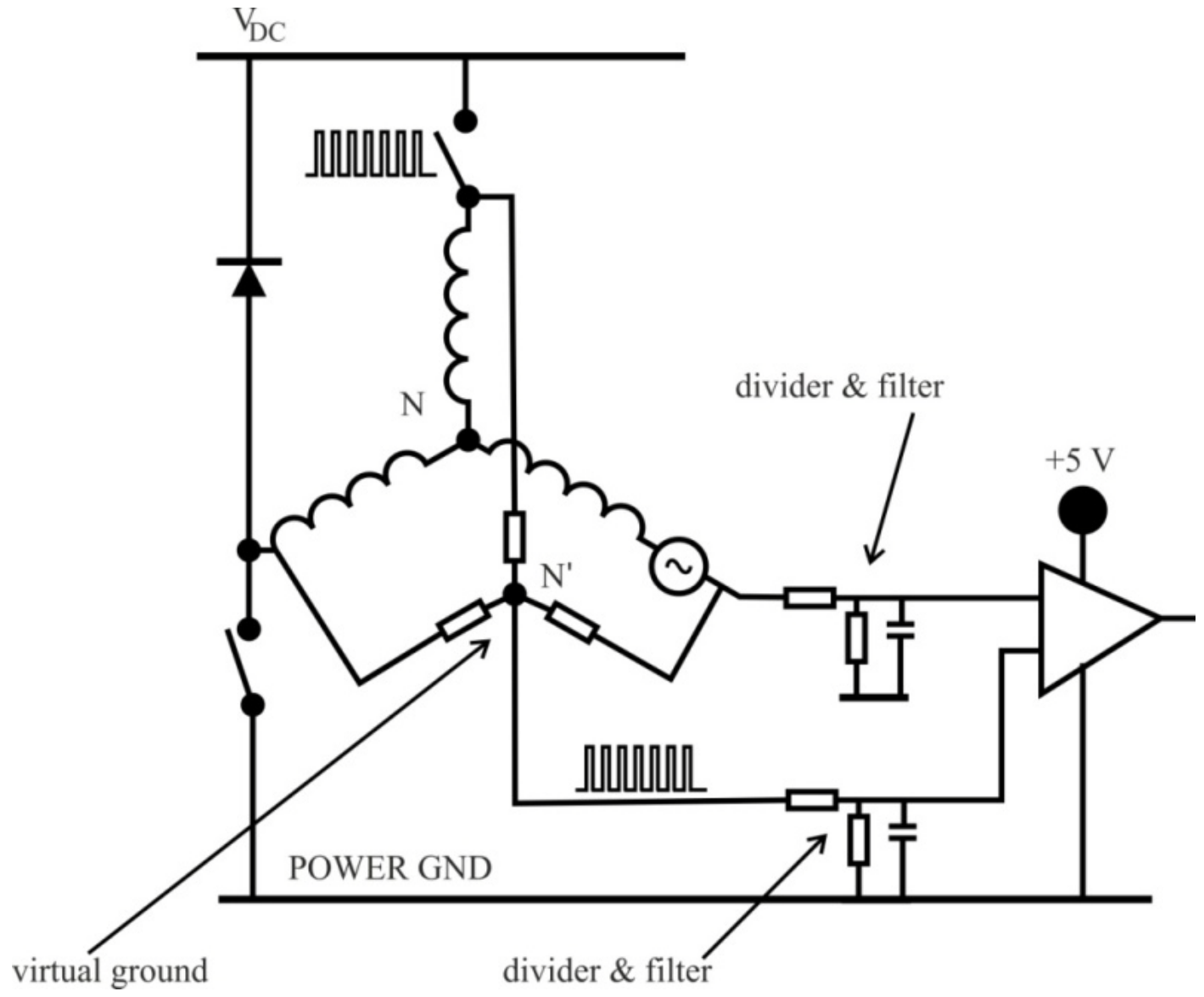

By means of eliminating the virtual neutral point when measuring back-EMF, a low amount of filtering is required, and the zero crossing of the back-EMF voltage of the floating phase can be obtained directly from the motor terminal voltage referred to ground by properly selecting the PWM and sensing strategy. Besides, the neutral point potential will be function of each phase's back-EMF, and will not be affected by any external driven voltage. Also, in this method the PWM signal is applied on high side switches only, and the back-EMF signal is synchronously detected during the PWM off time [22], as illustrated in Figure 14. The low side switches are only switched to commutate the phases of the motor. Then, the *true back-EMF can be detected during PWM off time* because the terminal voltage of the motor is directly proportional to the phase back EMF during this interval.

**Figure 14.** PWM applied to high side switches of a typical inverter for BLDC motors [22].

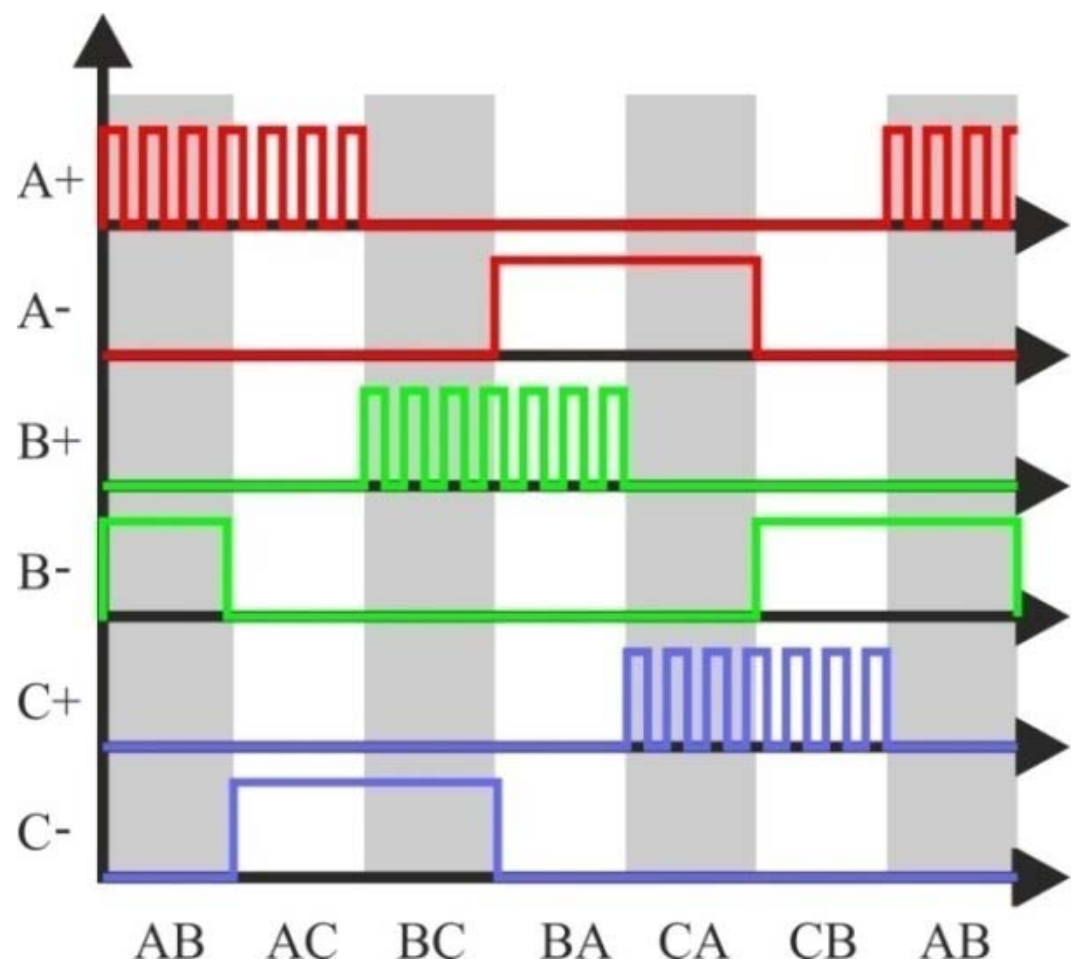

During the off time of the PWM, the terminal voltage of the floating phase is directly proportional to the back-EMF voltage without any superimposed switching noise, as Equation (10) indicates.

Besides, this terminal voltage is referenced to the ground instead of the floating neutral point, so the neutral point voltage information is not needed to detect the back-EMF zero crossing [24]:

$$V_N = \frac{e_C}{2} \Rightarrow V_C = e_C + V_N = \frac{3}{2} \cdot e_C \tag{10}$$

The resulting signal is not attenuated or filtered and it has a good signal/noise ratio, including a much wider speed range, which is beneficial at high speed operation. Then, this sensing technique has high sensitivity and it can be used in either high voltage and low voltage systems without effort to scale the voltage [22,24].

A good application for this sensorless drive system is in an automotive fuel-pump [38]. A dedicated sensorless BLDC controller can be incorporated in the fuel tank, implementing a back-EMF zero-crossing detection circuit as one of its peripherals, which simplifies the vehicle systems as well as reduces the overall system cost. This method has been also successfully applied to some home appliances for compressors, air blowers, and vacuum cleaner as well as engine cooling fan, and HVAC (Heating, Ventilating and Air Conditioning) blower motor applications.

3.5.3. Technique for low speed or low voltage applications

For low voltage applications, the voltage drop across the BJT's or MOSFET's will affect the performance. When the motor speed goes low, zero crossing is not evenly distributed. Besides, if the speed goes further low, the back-EMF amplitude becomes too low to detect [22].

There are basically two methods to correct the offset voltage of back-EMF signal [49]. One of them is to use *complementary PWM* as shown in Figure 15, which also reduces the conduction loss [22]. Another method is to eliminate the effect of diode voltage drop in order to add a constant voltage to compensate the effect of diode, and threshold voltage for avoiding the asymmetry in the distribution of zero crossing [24]. Then, in order to eliminate the non-zero voltage drop effect, a complementary PWM can be used, which will also reduce the power dissipation in the devices [49].

**Figure 15.** Complementary PWM algorithm [22].

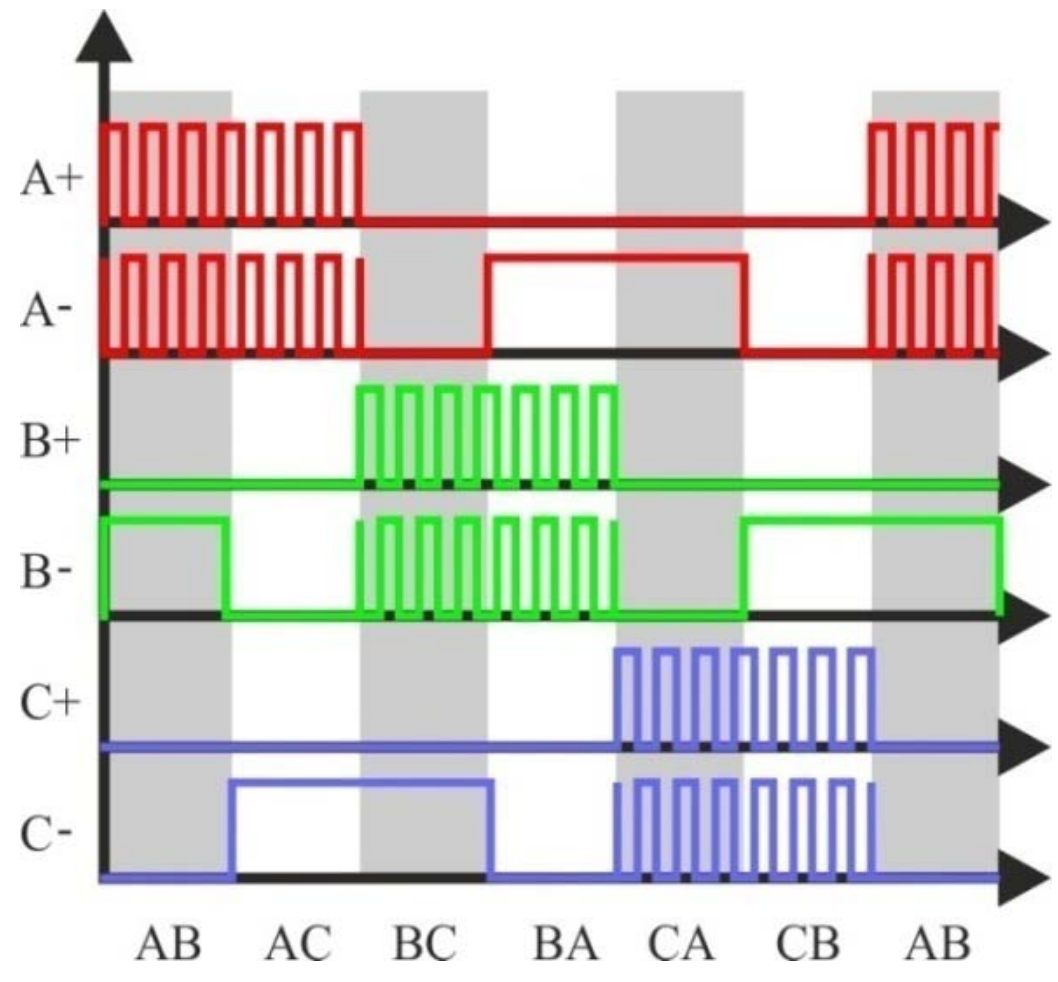

Assuming at a particular step, phase *A* and *B* are conducting current, and phase *C* is floating. The terminal voltage $V_C$ is sensed when the upper switch of the half bridge is turned off, and the current goes over the freewheeling diode *D*. During this freewheeling period, the terminal voltage $V_C$ is detected as phase *C* back-EMF. Then, the terminal voltage $V_C$ is shown in Equation (11), considering a low voltage MOSFET, in which $R_{DS(ON)}$ is very low and $V_{DS}$ can be ignored:

$$V_N = \frac{V_{DS} - V_D}{2} + \frac{e_C}{2} \Rightarrow V_C = e_C + V_N = \frac{3}{2} \cdot e_C - \frac{V_D}{2} \tag{11}$$

Therefore, the voltage drop on the diode will bias the terminal voltage of phase *C*. When the back-EMF $e_C$ is high enough at high speed, the effect of second term of Equation (11) is negligible [24]. However, at low speed especially during the start-up, the back-EMF itself is very small, and the second term will play a significant role. This *voltage offset will cause un-evenly distributed back-EMF zero-crossings*, which causes *unexpected commutation* and will affect the performance of the system. Also, because the back-EMF signal is too weak at low speed; an amplifier can be used as a pre-conditioning circuit for adjusting the offset and amplifying the signal near the zero-crossing [49]. Finally, the motor speed can be greatly expanded with the improvements explained before. For example, if a 48 V motor is used, the speed operation range can be from 50 rpm to 2,500 rpm [49].

#### 3.5.4. Technique for high speed or high voltage applications

One of the direct back-EMF sensing schemes analysed before could be implemented, for instance, in a hardware macro cell inside a microcontroller [49]. The three phase terminal voltages will feed into the microcontroller through resistors, which limit the injected current. When the PWM duty cycle is high, wrong zero-crossing detection occurs. This problem is caused by the large time constant of the current limit resistors. Additionally, there is some parasitic capacitance inside the microcontroller. Since the outside resistance is high enough, even though the capacitance is low, the effect of *RC* time constant will show up, and the falling edge of signal $V_C$ will be long.

As the back-EMF signal is sampled at the end of PWM off time, if the PWM duty cycle is high enough such that the off time is less than the falling edge time, the sampling result is not correct because the discharging period has not finished yet. To shorten the discharging time, the *RC* time constant should be reduced. A possibility is to use a smaller resistor in parallel with the resistor of the time constant, and a diode to block the charging current passing through the parallel resistor [49].

Since the back-EMF is sent directly to the microcontroller through a current limit resistor, for high voltage applications the resistance value is fairly high. This resistance with the parasitic capacitance in the circuit will cause too much delay and cause false detection. Using the improvements commented before, the sensorless system can be successfully used for 300 V/30,000 rpm high-speed air blower applications [50].

#### 3.5.5. Technique for small power applications

The basic concept for the development of this three-phase PWM technique is similar to that of synchronous rectifier for DC/DC converter. In a conventional approach [19], the switching losses are low in the inverter side at the cost of significantly high harmonic contents, which result in increase of loss in the motor side. For small power applications of BLDC drives, for instance fan and spindle applications, due to the use of battery and limited space for heat dissipation, reduction of power consumption becomes one of the main concerns in this PWM technique [22].

The high side power device is chopped in 1/6 fundamental period and duty ratio is derived from the speed reference or error of speed. Similar control signal is applied to the low-side power device with 180º shift. These control signals are applied to the other two phases with 120° shift. Under this condition, in the circuit of Figure 7 the current flows through the anti parallel diode of power device $T_{A-}$, and thereby resulting in significant conduction losses, which equal to the product of forward drop voltage of diode and load current. This significant heat loss is reduced in this technique [22]. The high-side power device is chopped in fundamental period and clamped to the positive DC link rail in the consecutive fundamental period [51], which is illustrated in Figure 16. As the high-side device is with chop control, the associated low-side power device is triggered by the inverse signal of chop control.

**Figure 16.** PWM waveform for low power PWM technique, where ωt[60°,120°] and TA+ is "off" [51].

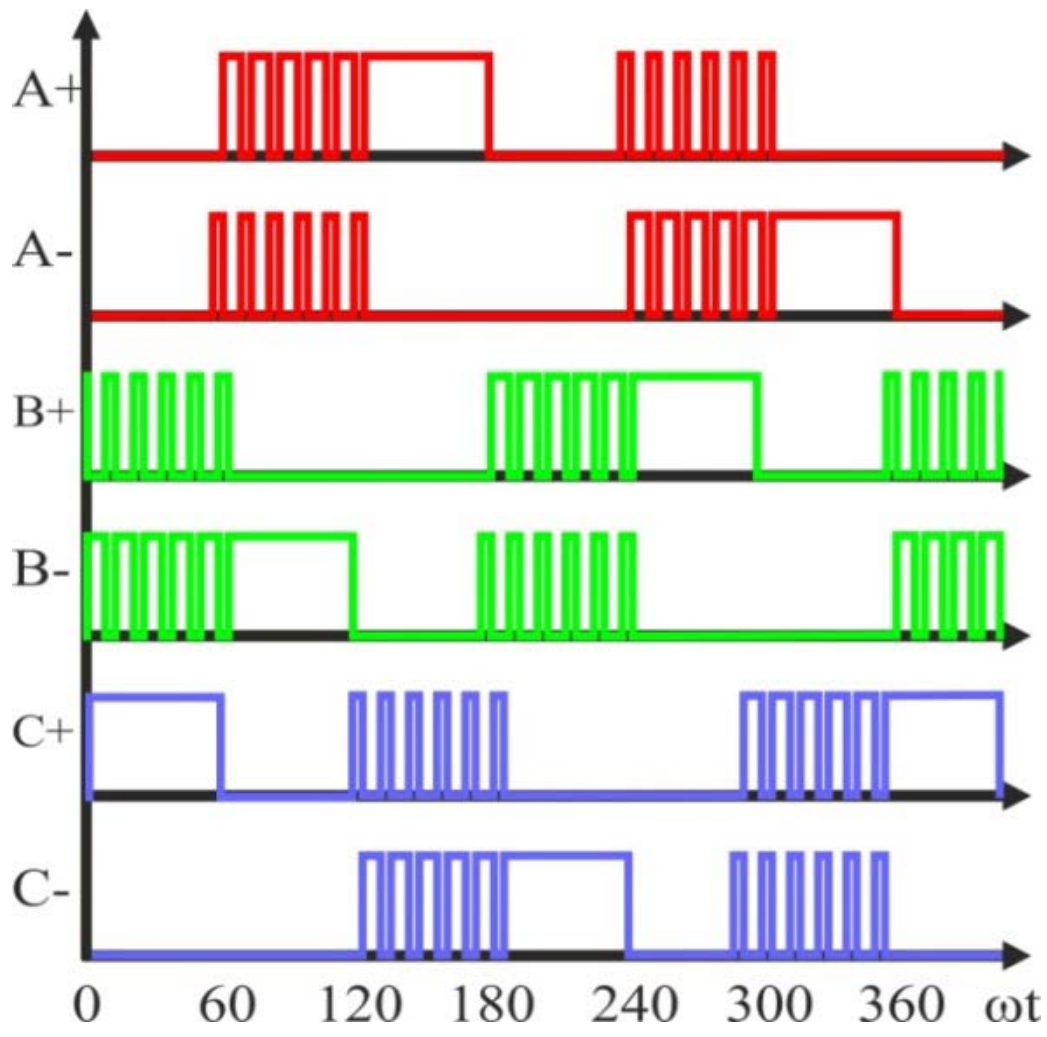

To highlight the feature of this PWM technique, the voltage drop caused by the turn-on resistance of power device and load current is significantly reduced as compared to the forward voltage drop of diode. Therefore, *the power consumption and the heat losses can be significantly reduced*.

A practical implementation of this method is in DVD spindle devices using FPGAs. Only terminal voltages of three phases are sampled and fed into the FPGA controller to calculate the commutation instants, and an external resistor and an eight-bit A/D converter are used to give the reference of duty ratio [48].

3.5.6. Direct current controlled PWM technique (hysteresis current control)

With the advantages of BLDC motor drives [2,41,52], it is possible to use a reduced converter configuration with advanced control techniques. Then one switch leg in the conventional six-switch converter, showed in Figure 7, is redundant to drive the three-phase BLDC motor, which results in the possibility of the four-switch configuration instead of six switches [21], as shown in Figure 17. This new drive consists of two switch legs and split capacitor bank, so two phases are connected to the switch legs and the other phase to the midpoint of DC-Link capacitors. But with this configuration, the limited voltages make very difficult to obtain 120° conducting profiles. This is the well-known problem *asymmetric voltage PWM* [21], which results in the 60° phase-shifted PWM strategy to generate three-phase balanced current profiles. Also, the conventional PWM schemes cannot be directly applied for the new drive configuration.

**Figure 17.** Four-switch converter for driving a three-phase BLDC motor [21].

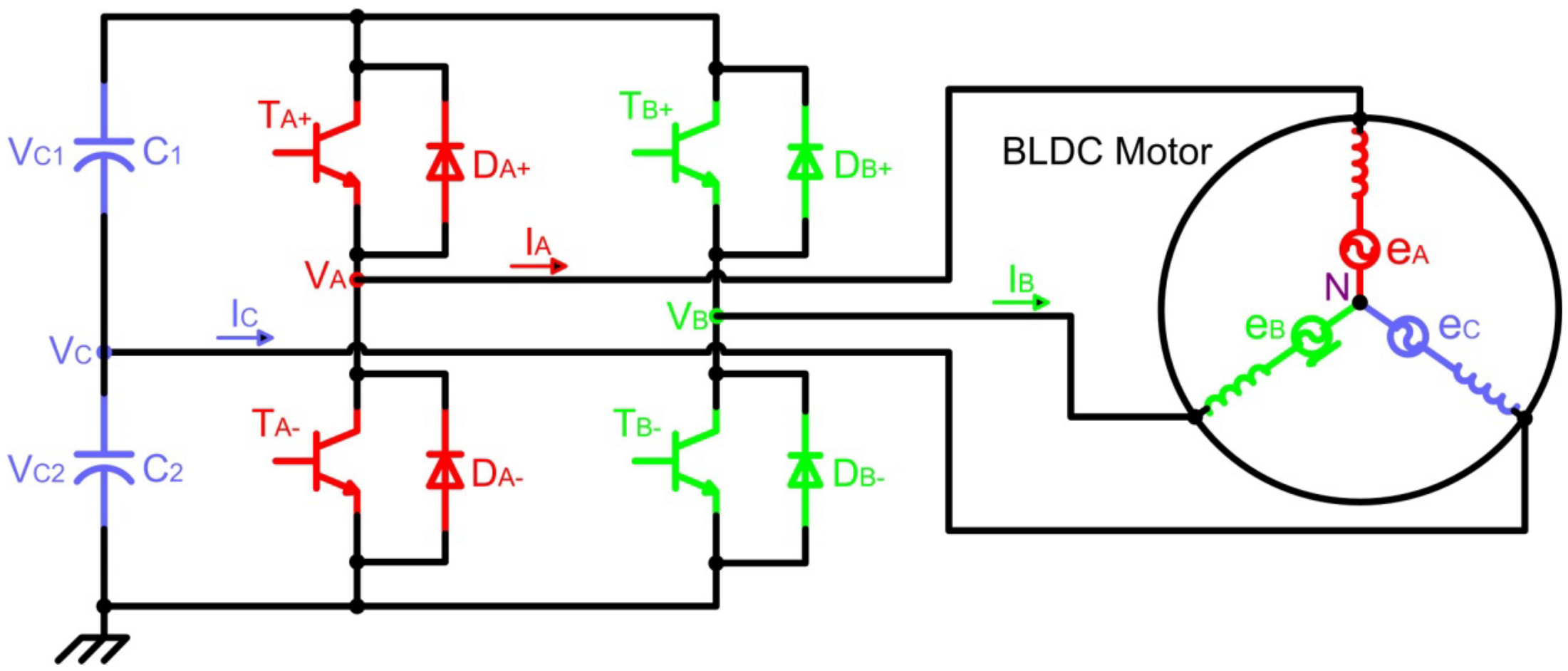

The direct current PWM control technique is based on the current controlled PWM method, instead of the voltage controlled PWM, which generates robust speed and torque responses and is simple to be implemented from the hardware and software points of view. Therefore, the four-switch three-phase BLDC motor drive, mainly applied to AC induction motor drives until now [53-55], could be a good alternative to the conventional configuration with respect to low-cost and high performance.

In a PWM control strategy for the four-switch three-phase BLDC motor drive, the three-phase currents always meet the condition of Equation (12):

$$I_C = -(I_A + I_B) \tag{12}$$

It means that control of the two-phase currents can guarantee the generation of the 120º conducting three-phase currents profiles. For completing this task, the two-phase currents are directly controlled using the *hysteresis current control* method by four switches [21].

## 4. Other Sensorless Techniques: Estimation and Model-Based Methods

It is convenient when designing feedback control systems, such as the motor position and speed, to assume initially that the entire state vector of the system to be controlled is available through measurement. If the entire state vector cannot be measured, as it is typical in most complex systems, the control law deduced cannot be implemented. Thus either a new approach that directly accounts for the nonavailabi1ity of the entire state vector must be devised, or a suitable approximation to the state vector that can be substituted into the control law must be determined [56]. In almost every situation, the development and use of an approximate state vector, which will substitute the unavailable state, is vastly simpler than a direct point of view of the system design.

Adopting this perspective, a control design problem can be split into two phases. The first phase is design of the control law assuming that the state vector is available, which may be based on optimization or other design techniques, and typically results in a control law without dynamics. The second phase is the design of a system that produces an approximation to the state vector. This system, called *observer* in a deterministic setting, has as its inputs the inputs and available outputs of the system whose state is to be approximated and has a state vector that is linearly related to the desired approximation [56]. Besides the simplicity of its design, the biggest advantage of using observers is that all of the states in the system model can be estimated including states that are hard to obtain by measurements [25]. In addition to their practical utility, observers offer an associated theory, which is intimately related to the fundamental linear system concepts of controllability, observability, dynamic response, and stability, and provides a simple setting in which all of these concepts interact.

To sum up, an observer provides a mathematical model of the brushless DC motor, which takes measured inputs of the actual system and produces estimated outputs. The error between the estimated outputs and measured quantities is fed back into the system model to correct the estimated values, such as the rotor position and speed, as would be the actually measured variables in a closed-loop system control [41]. Although most of the observer-based methods are used for PMAC motors, which have sinusoidal back-EMF and need continuous rotor position, for the BLDC motors, which require just six position points for one electrical cycle, the continuous position information from the observer is not necessary typically. But, for special purposes, such as flux weakening operation based on advanced angle control, the positions between commutation points are required [25].

### *4.1. Sliding-Mode Observer (SMO)*

For controlling BLDC motor, it is necessary to know an absolute position of the rotor, so an absolute encoder or resolver can be used for sensing the rotor position. But, these position sensors are expensive and require a special arrangement for mounting. Also, the state equation of BLDC motor is nonlinear, so it is difficult for the linear control theory to be applied and the stability of position and velocity estimation have not been clarified. To improve the mechanical robustness and to reduce the cost of the drive system, several estimation techniques eliminating the encoder or resolver can be applied [57]. Some relevant methods have been developed using the sliding-mode observer, which are briefly explained next.

In the Direct Torque Control method (DTC), the state equation of the BLDC motor is utilized to achieve a relationship between the angle of the stator current vector and the back-EMF vector angle, obtaining minimum error angle estimation and reducing the torque ripple in com-mutation regions. In this control method, the proper voltage vector is selected from a look-up table using the rotor flux vector position and torque error, which is led to the predefined hysteresis [58]. However, DTC methods based on hysteresis controllers have some serious drawbacks such as a high amount of torque and flux pulsations and variable switching frequency of the inverter [59]. Also, in the direct torque control of brushless DC motor, the stator flux linkage observation is needed, and the accuracy of the observed stator flux linkage is affected by the variation of stator resistance, electric interference, magnetic interference, measurement error and so on [60]. These drawbacks are solved with the DTC Space Vector Modulation (DTC-SVM) scheme, which uses a constant switching frequency. However, the DTC-SVM scheme needs a transformation from stationary reference frame to stator flux field orientation frame and vice versa, therefore it has a high computation time and could be an erroneous cumulative scheme [61]. Also, with the introduction of DTC technique and the advances of speed sensorless systems, the interest in stator resistance adaptation came to scene for an optimal performance of speed sensorless systems in low speed region [62].

Recently, and commented above, low speed operation with robustness against parameter variations remains an area of research for sensorless systems, taking into account that an accurate value of stator resistance is of utmost importance for its correct operation in low speed region. As in the upper speed range, the resistive voltage drop is small as compared with the stator voltage; hence the stator flux and speed estimation can be made with good accuracy. At low speeds the stator frequency is also low, but stator's voltage reduces almost in direct proportion and the resistive voltage drop maintains its order of magnitude and becomes significant. This greatly influences the estimation accuracy of the stator flux and hence the speed estimation. An estimation algorithm based on SMO in conjunction with Popov's hyper-stability theory can be used to calculate the speed and stator resistance independently, which can guarantee the global stability and the convergence of the estimated parameters [62].

The SMO is widely studied in the field of a motion control, and it can be applied to nonlinear systems, such as BLDC motors [63]. This technique applied to control systems encounters restrictions in practice, due to the high voltage values of the power supply needed and severe stress given to the static power converters. On the other hand, the sliding mode has been shown very efficient in the state estimation due to its salient features, *i.e.*, robustness to parameter variations and disturbances including the measurement noise. The use of sliding mode in state observer does not present physical restrictions relative to the convergence condition (the estimation error moves toward zero) and does not subject the system to undesirable chattering [57]. These problems can be alleviated using a **binary observer** with continuous inertial Coordinate-Operator Feedback [63].

### *4.2. Extended Kalman Filter (EKF)*

The extended Kalman filter algorithm is an optimal recursive estimation algorithm for nonlinear systems. It processes all available measurements regardless of their precision, to provide a quick and accurate estimate of the variables of interest, and also achieves a rapid convergence. This is done using the following factors: the knowledge of the system dynamics, statistical description of the system errors (noises, disturbances, *etc.*), and information about the initial conditions of the variables of interest. The algorithm is computationally intensive, thus an efficient formulation is needed rather than a straightforward implementation. Moreover, for a practical application of the filter in real time, different aspects of implementation have to be addressed, such as the computational requirements (processing time per filter cycle, required memory storage, *etc.*) and the computer constraints (cycle execution time, instruction set, arithmetic used, *etc.*) [64].

This method can be used to estimate the rotor position and speed. Motor state variables are estimated by means of measurements of stator line voltages and currents, and applying EKF next. During this process, voltage and current measuring signals are not filtered, and rotor position and speed can be estimated with sufficient accuracy in both steady state and dynamic operations [22]. Unlike the deterministic base of other studies, the model uncertainties and nonlinearities in motors are well suited to the stochastic nature of EKFs, as well as the persistency of excitation due to the system and measurement noises. This is the reason why the EKF has found wide application in speed-sensorless control, in spite of its computational complexity. However, with the developments in high performance processor technology, the computational burden and speed of EKF has ceased to be a problem [65].

The block diagram of the system for speed and rotor position estimation of a BLDC motor is shown in Figure 18. The system can be functionally divided in two basic parts: the speed control system and the estimation system. The first one consists of a power circuit (DC supply, inverter and motor) and control circuits, which perform three functions: current commutation, current control and speed control. The measured speed ($\omega_k$) and phase currents ($i_k$) as well as the estimated rotor position ($\hat{\theta}_{k/k}$) are used as feedback signals. The main blocks of the estimation algorithm are the EKF and the block for calculating average motor line voltages during sampling time. The average line voltages vector, defined on the basis of average line voltages in the *k*-sampling time ($u_k$), is calculated at the beginning of the sampling time by means of terminal voltages to neutral-point vector ($u_{Nk}$), the inverter transistors duty cycle ($\epsilon_k$), the inverter DC voltage ($U_0$), the estimated speed ($\hat{\omega}_{k/k}$), the rotor position ($\hat{\theta}_{k/k}$), and measured currents vector ($i_k$) [66].

**Figure 18.** System configuration for speed and rotor position estimation of a BLDCM [66].

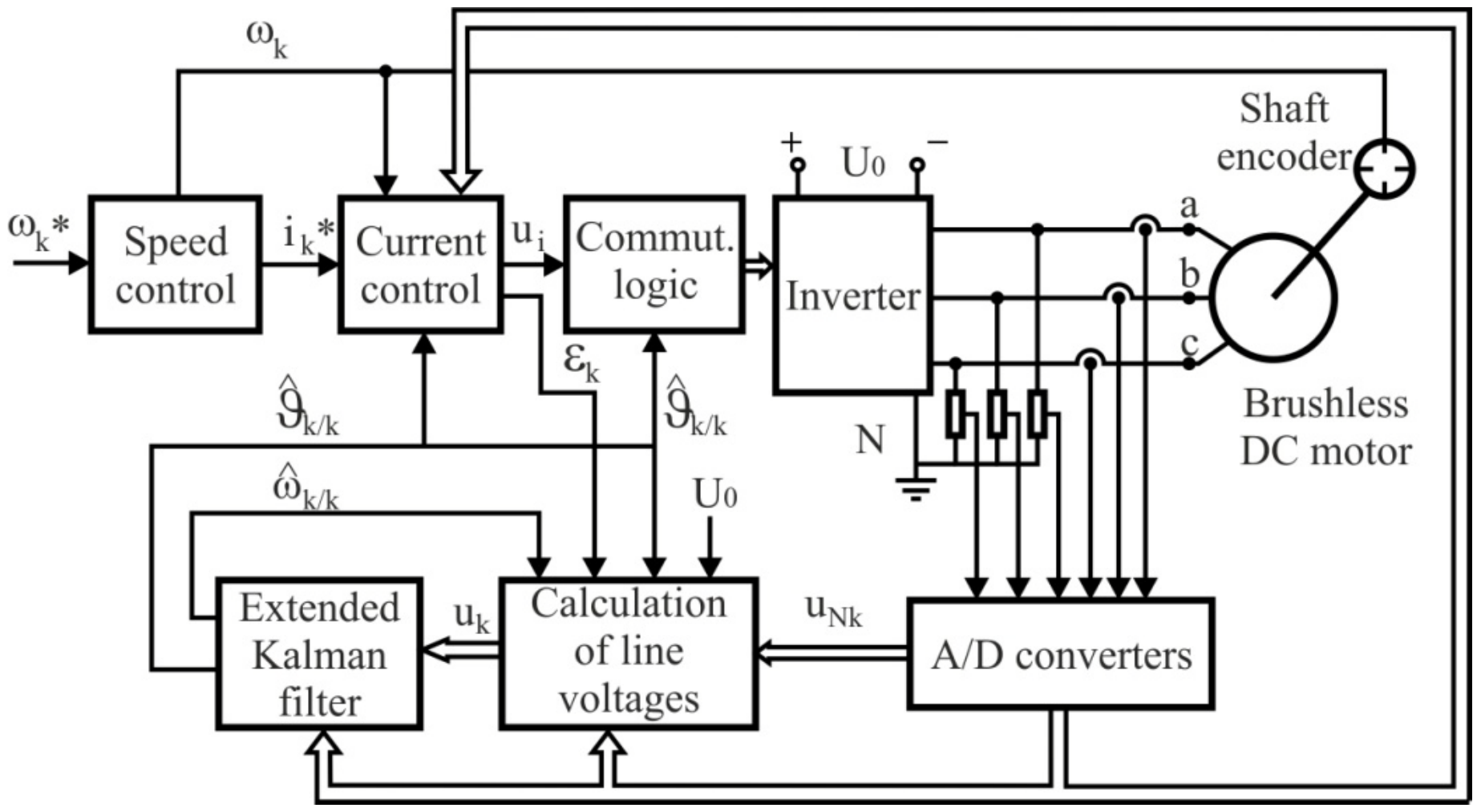

Among recent speed-sensorless studies using EKF based estimation, the simultaneous estimation of the rotor angular velocity, the rotor flux and the stator resistances, via a Kalman filter in combination with the model reference adaptive system (MRAS), have been performed, but are sensitive to variations in the stator and rotor resistances. Some innovative techniques have been currently developed, such as the *Bi Input-EKF* (BI-EKF). This method utilizes a single EKF algorithm with the consecutive execution of two different inputs, which are calculated from the two extended models based on the rotor and stator resistance estimation, respectively. These two different inputs are used for the rotor flux based speed control both in the transient and steady-state over a wide speed range. Also, the load torque is estimated, including viscous friction term, rotor angular velocity, rotor flux, and stator current components without the need for signal injection [65].

### *4.3. Model Reference Adaptive System (MRAS)*

In some cases, the stator and rotor resistance estimation is not applicable when the speed-sensorless control system is in transient state, such as operation under largely varying load torque and/or changes in the speed command. In other cases, the rotor time constant via high frequency signal injection, the stator resistance and the rotor angular velocity can be estimated by using MRAS. However, the stator resistance estimation is turned on for short time intervals when the rotor angular velocity estimation has reached its steady-state; that is, both the stator resistance and rotor angular velocity estimations are performed interchangeably [65].

The model reference adaptive system, developed using Popov's stability criterion, is one of many promising techniques employed in adaptive control for estimating the speed and stator resistance [62]. Among various types of adaptive system configuration, MRAS is important since it leads to a relatively easy-to-implement system with a fast adaptation for a wide range of applications. The basic principle is illustrated in Figure 19, called *parallel MRAS*. The dynamic models are represented by the block "Reference Model", which is the actual system (for example, the motor, containing all unknown parameters, *i.e.*, motor speed, stator and rotor resistances) and the block "Adjustable Model", which has the same structure as the reference one (*i.e.*, motor, but with the adjustable or estimated parameters, instead of the unknown ones). An error vector $\epsilon$ is derived using the difference between the outputs of two dynamic models and is driven to zero through an adaptation law. As a result, the estimated parameter vector will converge to its true value *X*. One of the most noted advantages of this type of adaptive system is its high speed of adaptation. This is due to the fact that a measurement of the difference between the outputs of the reference model and adjustable model is obtained directly by comparison of the states (or outputs) of the reference model with those of the adjustable model system [67]. It is remarkable that the error signal may be formulated with flux (F-MRAS), back-EMF (E-MRAS), reactive power (Q-MRAS) and active power (P-MRAS) [68].

**Figure 19.** Basic configuration of a MRAS [67].

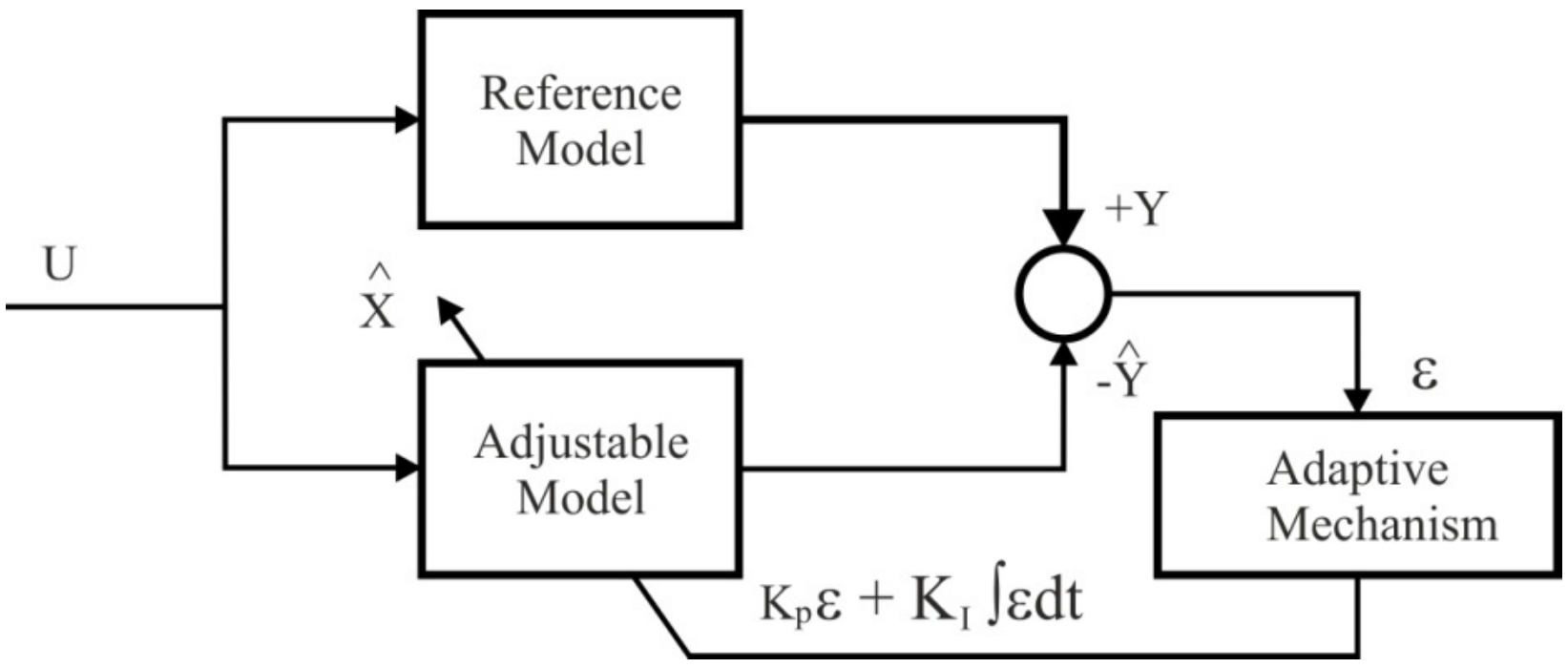

For instance, a MRAS with instantaneous *reactive power* can be used for speed estimation of sensorless vector controlled motor drive. This MRAS converts a vector quantity (*i.e.*, current vector) into a scalar quantity using the concept of reactive power, and the reference model utilizes measured current vector. Also, the adjustable model uses the estimated stator current vector, and the current, estimated through the machine state equations, is configured in terms of reactive power [68]. An *active power* MRAS based scheme can also be used for rotor resistance identification, whose estimation is effective in wider range of variations and could be applied in real time field-oriented control (FOC) [67].

### *4.4. Adaptive Observers*

The interest in stator resistance adaptation came to scene much recently, with the advances of speed sensorless systems and with the introduction of DTC technique. An accurate value of the stator resistance is of crucial importance for correct operation of a sensorless drive in the low speed region, since any mismatch between the actual value and the set value used within the model of speed estimation may lead not only to a substantial speed estimation error but also to instability as well. Therefore, to develop online stator resistance identification schemes are of utmost importance for accurate speed estimation in the low speed region. These estimators often use an adaptive mechanism to update the value of stator resistance. Some of the most relevant are MRAS, explained previously, and adaptive full-order flux observers (AFFO) [62].

#### 4.4.1. Adaptive Full-order Flux Observer (AFFO)

The AFFO scheme has been developed using Lyapunov's stability criterion and allows estimating the rotor speed and stator resistance simultaneously. Using this observer, the estimated quantities converge to their real values if the persistency of excitation condition is also satisfied. Correct estimation of rotor flux space vector and rotor speed is therefore possible through this observer according to the stator and rotor resistances online adaptation [59]. Jointly with MRAS, AFFO is not computationally intensive, but with a non-zero gain matrix may become unstable. In such methods, the stator resistance adaptation mechanism is determined with the difference between the measured and observed stator currents [62]. With a maximum torque per ampere (MTPA) strategy, based on slip frequency (inverse of the rotor time constant in the rotor flux oriented reference frame) adjustment, the stator current amplitude can be minimized for each value of motor speed [59].

Apart from the variation of the stator resistance with temperature, other parameters in the AFFO will change during operation as well, such as the rotor resistance due to temperature changes, which will have an important influence on the speed accuracy of the adaptive observer. The stator and rotor self-inductance and magnetizing inductance vary due to magnetic saturation, being it possible to use a nonlinear magnetic model. In steady state, it is known that a misestimation of the rotor resistance provides correct estimations of the stator and rotor flux, but results in a misestimation of the speed [69].

This observer can be also applied in both rotor and stator-flux-vector-controlled drives, and in DTC drives. Even, it can be used in combination with the *Discrete-Time Direct Torque Control method* (DT-DTC), whose advantage over the classical DTC method is the torque-ripple-free operation in the whole speed range [69].

4.4.2. Adaptive Pseudoreduced-Order Flux Observer (APFO)

The field-oriented control (FOC) concept is widely used for high performance control of AC motors. In the case of induction or brushless drive systems, the FOC requires a speed sensor, such as shaft-mounted tachogenerator, resolver, or digital shaft position encoder, for obtaining speed information, which spoils the ruggedness and simplicity of the motor. Therefore, several speed observers have been developed to solve this problem, such as AFFO for estimating the rotor flux and speed, whose computation process is simple but it is an unstable system in some cases and unsuitable for practical applications (large speed errors may appear under heavy loads and steady-state speed disturbances may occur under light loads) [70]. Besides, the MRAS technique is applied to sensorless speed control using the field oriented reference frame. While this method is simple and always stable, its performance is poor in the low speed range, where open-loop integration may lead to instability due to stator resistance misestimation, and it is very sensitive to the offset of the voltage sensor and the stator resistance variation due to temperature [71].

An improved scheme is the Adaptive Pseudoreduced-order Flux observer (APFO), which uses the Lyapunov method and consumes less computational time with a better speed response than the AFFO over a wide speed range [70]. An innovative idea applied in this observer is a constant gain matrix, which ensures that the poles are not related to the motor speed. Then the observer poles are invariant to any speed variations, so the drive can operate at high speeds and have fast response [72].

*4.5. Artificial Neural Networks (ANN)*

A neural network or artificial neural network (ANN) is the interconnection of artificial neurons that tends to simulate the nervous system of a human brain. Each signal coming into a neuron along a dendrite passes through a synaptic junction. The adjustment of the impedance or conductance of the synaptic gap leads to "memory" or "learning" process in the case of brain, which is similarly required in ANN. The model of an artificial neuron that closely matches a biological neuron is given by an op-amp summer-like configuration [73].

In recent years, the use of artificial neural networks for identification and control of nonlinear dynamic systems in power electronics and motor drives have been extended, as they are capable of approximating wide range of nonlinear functions to a high degree of accuracy. If a motor drive is considered, where the essential sensor signals relating to the state of the system are fed to a neural network, the network output can interpret the "health" of the system for monitoring purposes and control. ANNs can be also used for realization of current-regulated PWM inverters, in which the network receives the phase current error signals and generates the PWM logic signals for driving the inverter devices [74].

Referring to motors, ANNs can be used to estimate rotor flux, unit vector, and torque in vector-controlled drives. The network has to be trained with a very large number of simulation data sets, so DSP-based estimators and ANN-based estimators perform comparably [73]. The capability of a neural network can be deployed to have online estimators to address the situation of similar disturbances in both stator and rotor resistances simultaneously. In this situation, the resistance observer can be realized with a recurrent neural network trained using the standard back-propagation learning algorithm. The block diagram of a rotor flux oriented induction motor (IM) drive together with both stator and rotor resistance identifications are shown in Figure 20, which also applies to BLDC motors. A rotor flux oriented vector controller commands the motor, and the voltage model fluxes are estimated from the measured stator voltages and currents using a programmable cascaded low-pass filter (PCLPF). The stator voltages are PWM voltages, which are filtered, and only the sinusoidal voltages are taken into the PCLPF flux estimators. The RRE and SRE blocks estimate the rotor and the stator resistances, respectively, taking into consideration that the stator resistance estimation depends on the rotor resistance. The rotor flux linkages are compensated by the RRE and hence the estimation error in the stator resistance is avoided [75].

**Figure 20.** Block diagram of the indirect vector controlled induction motor drive with online stator and rotor resistance tracking [75].

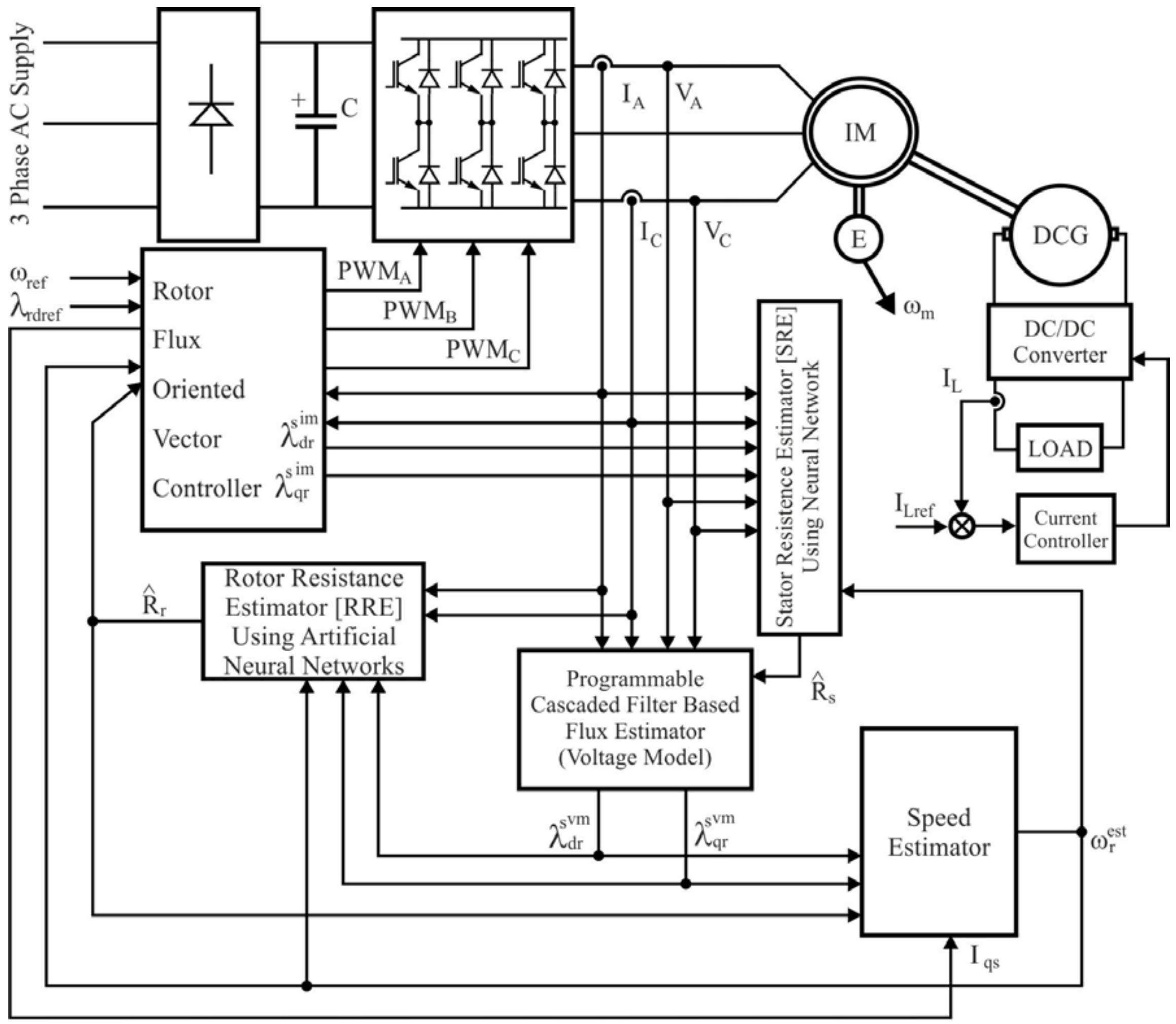

An improvement is the *Fuzzy Neural Networks* (FNN), which are systems that apply neural networks principles to fuzzy reasoning and can be based on rule or relational approaches. Basically, they emulate fuzzy logic controllers, with the advantages that it can automatically identify fuzzy rules and tune membership functions for a problem. FNNs can also identify nonlinear systems to the same precision as conventional fuzzy modelling methods, such as the development of stator flux-oriented vector-controlled drive [73].

## 5. Implementation of back-EMF Control Techniques

In the design and development of motor controllers based on back-EMF techniques, it is necessary to analyse the implementation options using electronic processors or ASIC. This is done by taking into account that all back-EMF sensing methods need to start the motor from standstill using an open-loop starting technique. All these issues are studied next.

### 5.1. Comparison of methods' feasibility

Assuming a three-phase BLDC motor as a reference model, a six-step commutation with 120º conduction time allows for current to flow in only two phases at any one time, which leaves the third phase available for sensing back-EMF. *Originally the method of sensing back-EMF* was proposed in order to build a virtual neutral point, which, in theory, will be at the same potential as the center of the wound motor. However, when using a chopping drive, the PWM signal is superimposed on the neutral voltage, inducing a large amount of electrical noise on the sensed signal [49]. To reduce the switching noise, the *back-EMF integration* [23] and *third harmonic voltage integration* [40] were introduced. The integration approach has the advantage of reduced switching noise sensitivity. However, it still has an accuracy problem at low speed. An indirect sensing of zero crossing of phase back-EMF by *detecting the conducting state of free-wheeling diodes* in the unexcited phase is complicated and costly, while its low speed operation is a problem [19]. The *back-EMF zero-crossing detection* (terminal voltage sensing) method, which does not require the motor neutral voltage, permits the extraction of the true back-EMF directly from terminal voltage by properly choosing the PWM and sensing strategy [24,76]. As a result, this sensorless BLDC driver can provide a much wider speed range from start-up to full speed than the conventional approaches.

The *third harmonic* based method is one of most relevant back-EMF sensing schemes. It has a wider speed range and smaller phase delay than the *terminal voltage sensing* method [77]. However, at low speed, the integration process can cause a serious position error, as noise and offset error from sensing can be accumulated for a relatively long period of time [25]. At lower speeds, detection of both the third harmonic and the zero-crossing of the phase voltage become difficult due to the lower signal levels. In comparison, the conventional back-EMF control scheme is able to drive the motor from 6,000 rpm to about 1,000 rpm, but the third harmonic control scheme is capable to operate the motor from rated speed (6,000 rpm) down to about 100 rpm. This does not introduce as much phase delay as the zero-crossing method and requires less filtering [41]. Then, the efficiency drop is more accentuated for the terminal voltage sensing scheme, because the delay introduced by the low pass filter decreases with the motor speed. This phase delay introduced by the filter is responsible for the loss of field orientation and loss of the quadrature condition between rotor flux and stator current. The immediate consequence is the reduction of the torque per current ratio of the motor, which implies in larger copper losses [40]. Also, the third harmonic back-EMF method is applicable for the operation in flux weakening mode, and the methods based on zero-crossing of the back-EMF are simple. However, it is only applicable under normal operating conditions (commutation advance or current decay in free-wheeling diodes lower than 30 electrical degrees) [22].

Due to most popular and practical sensorless drive methods for BLDC motors relying on speed-dependent back-EMF, and the back-EMF is zero or undetectably small at standstill and low speeds, it is not possible to use the back-EMF sensing methods in the low speed range. Then, *in all back-EMF-based sensorless techniques, the low-speed performance is limited, and an open-loop starting strategy is required* [18]. However, other control schemes such as the *PWM method eliminating the virtual neutral point* can provide a much wider speed range from start-up to full speed than the conventional schemes. Besides, this method has high sensitivity; it is good at high-speed operation due to a lack of unwanted delays in the zero-crossing detection. It can be easily used in both high-voltage or low-voltage systems, and it has a faster motor start-up because of the precise back-EMF zero-crossing detection without attenuation. Also, it is simple and easy to implement [38].

### *5.2. Open-Loop Starting*

The back-EMF detection methods can not be applied well when the motor is at a standstill or low speed, since back-EMF is zero. A starting procedure is needed to start the motor from standstill [20]. The open-loop starting is accomplished by providing a rotating stator field which increases gradually in magnitude and/or frequency. Once the rotor field begins to become attracted to the stator field enough to overcome friction and inertia, the rotor begins to turn and the motor acts as a permanent magnet synchronous machine with the disadvantage that the initial rotor movement direction is not predictable. When the stator field becomes just strong enough, the rotor could move in either direction. If the speed of the stator field is slow enough and the load torque demanded does not exceed the pull-out torque the motor will operate synchronously in the desired direction [41]. The change over from open-loop to sensorless method is made when sufficient back-EMF is generated, so that the sensorless method should start generating the switching instants of all transistors [7].

Taking all this into account, the procedure starts by exciting two arbitrary phases for a preset time (for example, 0.5 s). At the end of the preset time, the open-loop commutation advancing the switching pattern by 120° is done, and then, the polarity of the motor line current is altered. Then, the rotor turns to the direction corresponding to the exited phases as is shown in Figure 21a. Next, the commutation signal that advances the switching pattern by 120° is given, as Figure 21b indicates, and the open-loop commutation is immediately switched to the sensorless drive. After the next commutation the position sensorless drive is attained (Figure 21c), and the motor line current indicates that satisfactory sensorless commutations are performed by the position-detecting method [19].

**Figure 21.** Open-loop starting procedure [19].

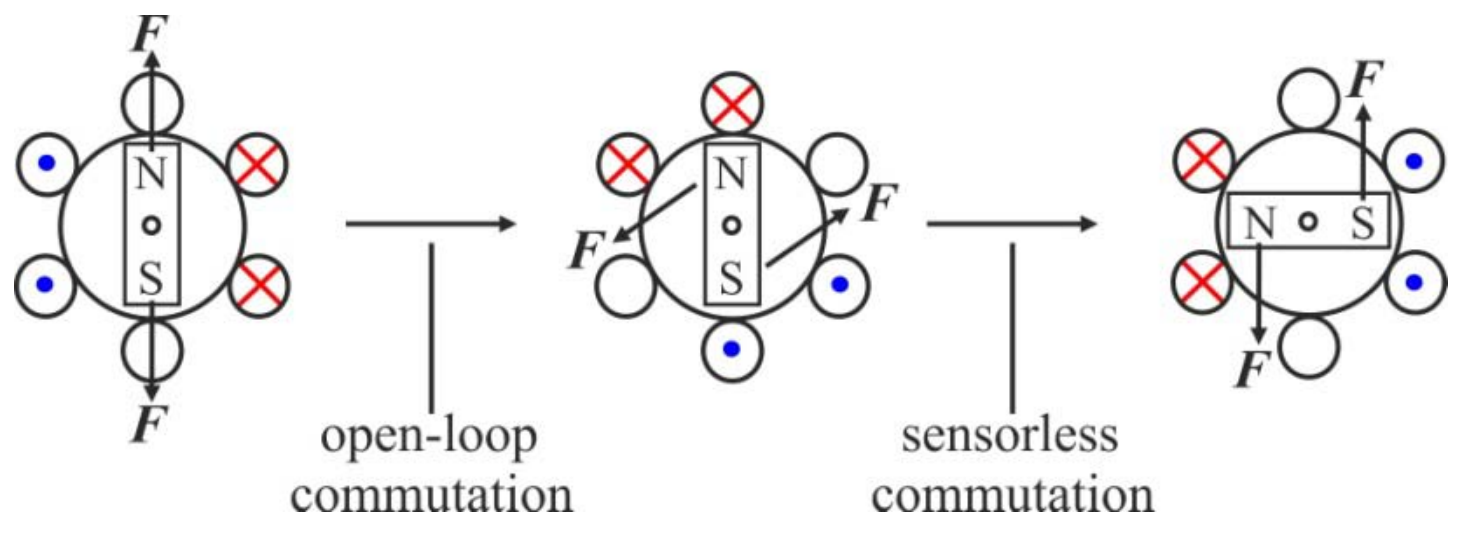

This method is simple but the reliability is affected by the load and it may cause temporarily reverse rotation of the rotor during the start-up. This is not allowed in some applications [20], such as disk drives, which strictly require unidirectional motion. However, it may be satisfactory in others such as pump and fan drives. Another problem exists if the stator field is rotating at too great a speed when the rotor field picks up. This causes the rotor to oscillate, which requires the stator field to decrease in frequency to allow starting.

The stator iron of the BLDC motor has non-linear magnetic saturation characteristic, which is the basis for determining the initial position of the rotor. In order to overcome the drawbacks mentioned above, the rotor position detecting and speed up methods based on saturation effect of the stator iron can be applied [78], such as the *short pulse sensing* technique. This scheme adopts a voltage pulse train composed of the successive short and long pulses to generate positive torque to speed up the motor, and it does not bring any reverse rotation and vibration during the start-up process [20]. The response speed of the stator current [79], and the response peak value of the current of the stator winding can be used to detect the rotor position [80,81].

### *5.3. Development of Controllers Using Electronic Processors and Specific Circuits*

The development and implementation of different control methods for BLDC motors can be carried out by using commercial processors and specific circuits. Some of the most common electronic devices applied to back-EMF sensing techniques are explained through different implementation models and examples.

#### 5.3.1. Digital Signal Processors (DSP)

Recently, the microprocessor technology has shown incredible improvements and the operating speed of DSPs have gotten faster and faster. Complicated control algorithms can now be easily implemented in a DSP with high sampling and calculation frequency. In addition, optimal performance of BLDC machine drive cannot be achieved with a control method based on the rotor position obtained from a position sensor such as encoders or Hall sensors, which are insensitive to parameter variation [25]. The combination of advanced DSP technology and sophisticated control will yield a better sensorless BLDC drive that has higher performance than even conventional motor drives with position Hall sensors in the near future. *In position or speed sensorless control* of drives the goal is to eliminate the sensors by powerful DSP based computation using the machine terminal voltage and current signals [82].

The Field Oriented Control (FOC) technology has been developed as a way to achieve high dynamic performance drive systems. The Universal Field Oriented (UFO) controller was introduced as a generalization for decoupling techniques for flux and torque vectors. The controller basically allows for the selection of what reference frame to use for the decoupling, depending on the operating characteristics of the drive system. The third harmonic voltage component was used solely to estimate the air gap flux of the motor, and a speed sensor was used in order to implement the complete control [83]. The acquisitions of the third harmonic voltage and rotor slot ripple signals are accomplished via the analog interface presented in Figure 22.

**Figure 22.** DSP and analog interface used to detect the stator phase voltages sum and the rotor slot harmonic component [83].

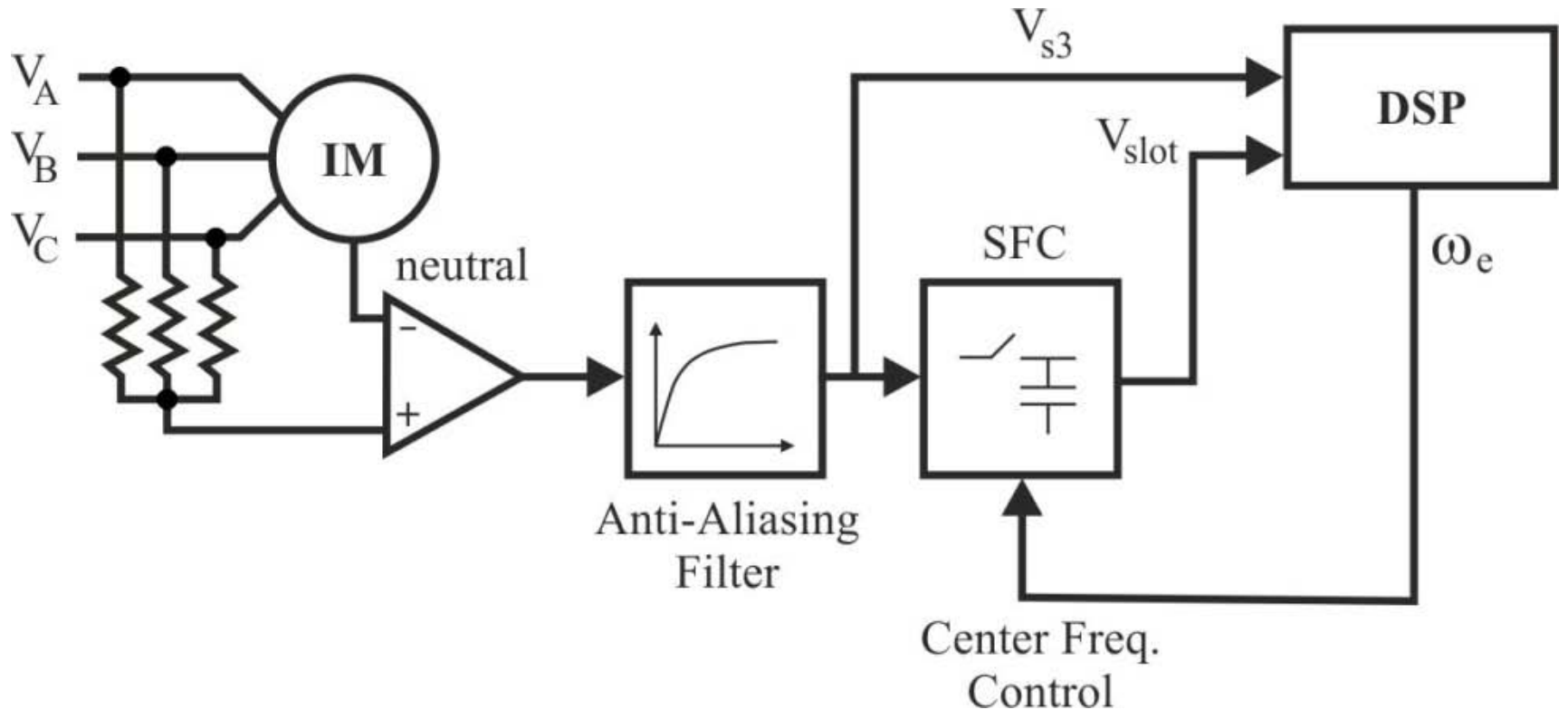

The block diagram of Figure 22 includes a DSP Texas Instruments TMS32010 [84] used to implement the control algorithms. The air gap flux and the rotor speed are detected by processing the signal obtained from the summation of the stator phase voltages, $V_{s3}$. The DSP performs the integration of the third harmonic voltage signal to derive the third harmonic flux. In order to detect the rotor speed the signal $V_{s3}$ is processed by a switched capacitor band-pass filter (SCF), whose central frequency can be tuned over a wide range (from about 20 Hz to 4 kHz). *The output of the filter is a variable amplitude sinusoidal wave. This wave has the same frequency as the rotor slot ripple* [83]. Two options are to detect the frequency of the SCF output signal: a Phase Locked Loop or a frequency to voltage converter (FVC).

5.3.2. Field Programmable Gate Arrays (FPGA)

A remarkable application of DSPs or FPGAs is the sensorless control for high speed applications based on the execution of PWM control schemes, which are classified as unipolar [38] and bipolar methods [24,51]. Depending on the PWM method used the control scheme may cause a commutation delay in high speed applications since the PWM switching and the inverter commutation cannot be done independently. If the commutation instant is synchronized with the end of the PWM switching period ideal commutation occurs with any delay. But since the *commutating instant depends on the rotor position* it does not generally coincide with the end of PWM period and undesirable commutation delay is produced. This problem can be overcome by *controlling the voltage and frequency independently by DC link voltage control scheme.* This control can be implemented using a DSP or FPGA based high speed sensorless control configuration [22].

Typical high speed applications in which PWM techniques can be applied are digital video disk (DVD) spindle systems, which can be implemented using a FPGA, such as the Altera Flex EPF6024AQC240-3 [85]. The controller includes two main parts: the PWM generation circuit and the power device control circuit. Figure 23 shows the system, which consists of a FPGA, a BLDC motor, and the associated reference and sensing circuits [51]. Only terminal voltages of three phases are sampled and fed into the FPGA controller to calculate the commutation instants. This system contributes to significant reduction of conduction losses and power consumption, which is quite essential for small power BLDCM drives powered by battery and/or with limited dissipation space.

**Figure 23.** Block diagram of the FPGA-based drive for DVD spindle applications [51].

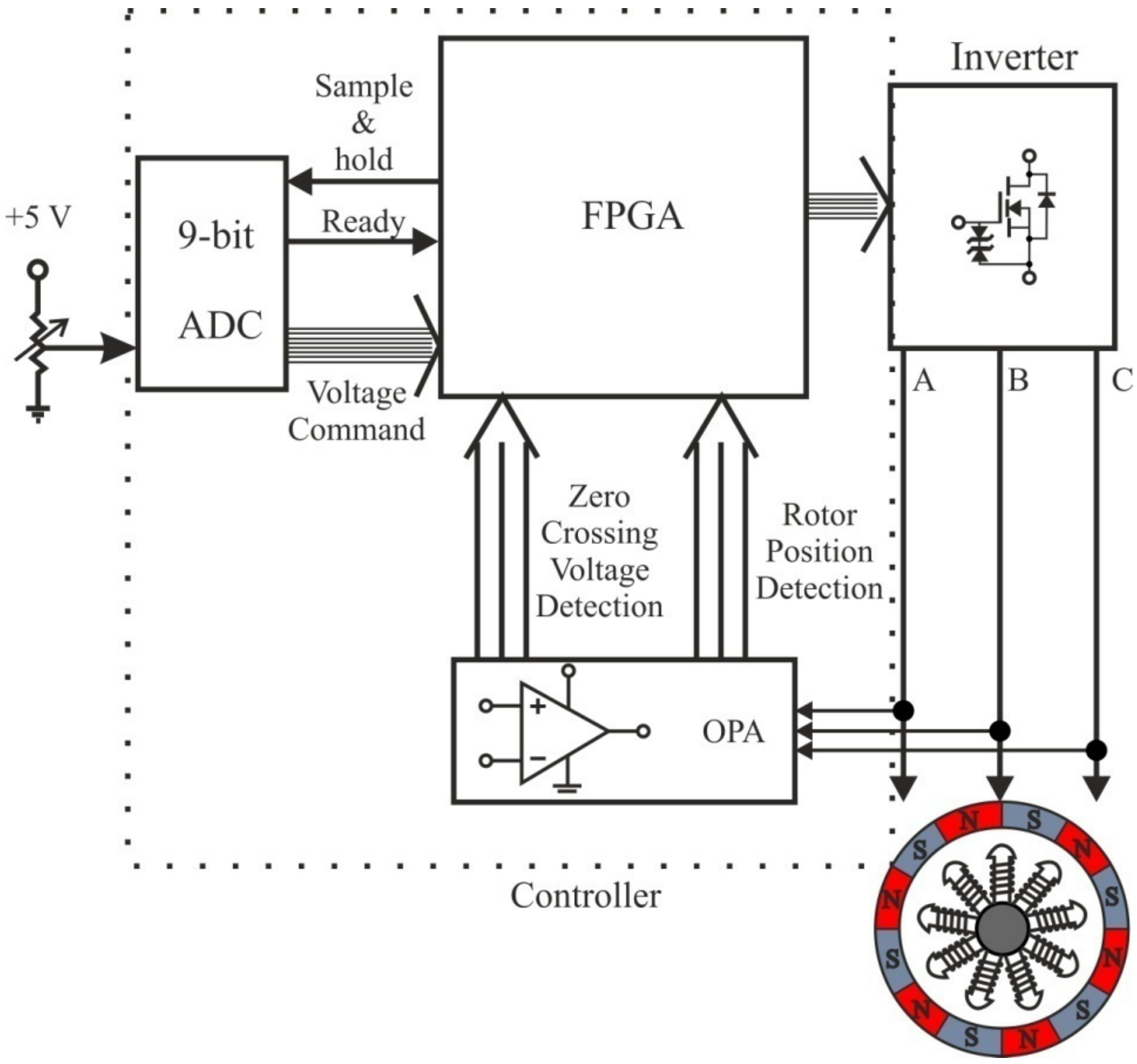

5.3.3. Microprocessors (MP)

A low-cost sensorless control scheme for BLDC motors can be implemented if rotor position information is derived by filtering only one motor-terminal-voltage, which leads to significant reduction in components count of the sensing circuit. As indicated in Figure 9, only two of the three state-windings are excited at a time, and the third phase is open during the transition periods between the positive and negative flat segments of the back-EMF [22]. Therefore, each of the motor terminal voltages contains the back-EMF information that can be used to derive the commutation instants.

Cost saving is further increased by coupling the position sensing circuit with a single-chip microprocessor or DSP for speed control. Figure 24 shows a block diagram of the position detection circuit based on sensing all three motor terminal voltages for a BLDC motor. Each of the motor terminal voltages, referred to as the negative DC bus rail $V_A$', $V_B$' and $V_C$' are fed into a filter through a voltage divider of a resistor network. This removes the DC component and high frequency contents that result from the PWM operation. The phase information is extracted from the back-EMF. The correction is based on measuring the elapsed time between the last two zero-crossing instants and converting it to frequency. This operation is achieved when the filtered voltage, $V_A$", is passed to a comparator to detect these zero-crossing instants, which are further sent to a microprocessor for phase-delay correction and generation of commutation signals. The microprocessor produces gate control signals for the inverter and may perform closed speed control with the motor speed information measured by the frequency of the detected signals [6].

**Figure 24.** Block diagram of the FPGA-based drive for DVD spindle applications [6].

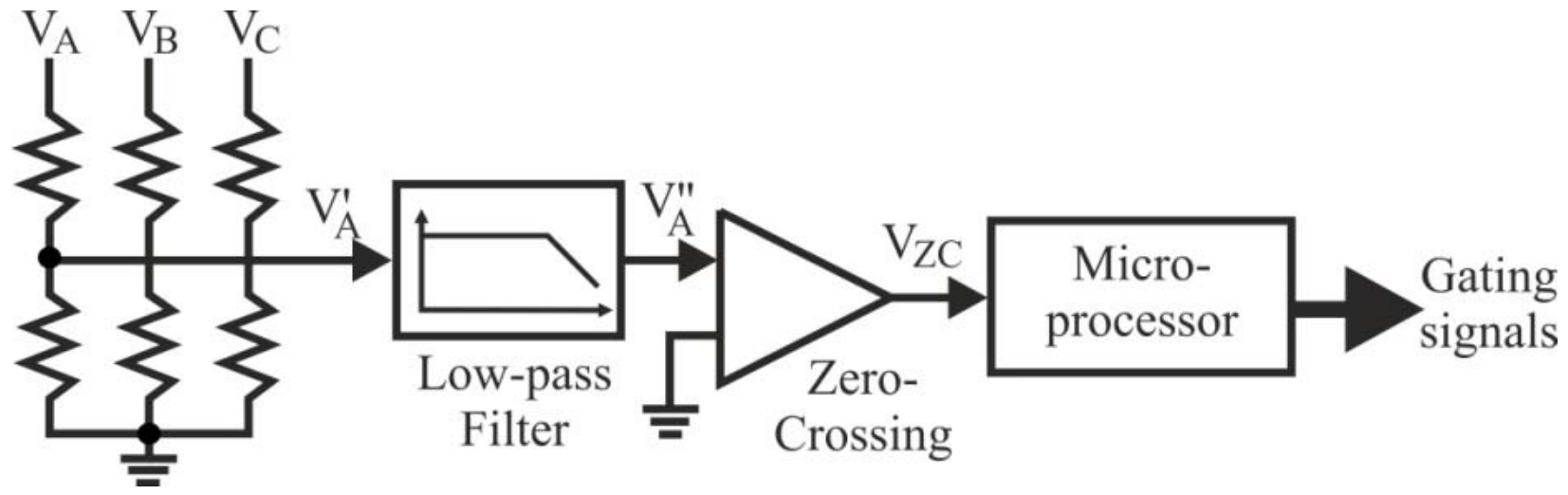

5.3.4. Microcontrollers (MC)

In recent years, with the development of mixed-signal integrated circuit technology, many system-on-chip (SOC) devices have become available. High-throughput microcontrollers with imbedded programmable memory as well as other precision analog and digital peripherals can be incorporated into a single integrated circuit. SOC devices have many advantages, including lower overall system cost and reduced board space, as well as superior system performance and reliability. Taking all these features into account, a dedicated sensorless BLDC controller implementing a back-EMF zero-crossing detection circuit as one of its peripherals, can be developed [24].

A back-EMF sensing method that requires neither a manufactured neutral voltage nor a great amount of filtering can be implemented using a microcontroller reducing the total system cost [76]. A usual microcontroller model is the ST72141 (STMicroelectronics) [50] which integrates the analog detection circuit and other motor control peripherals with a standard microcore. In this method, the true back-EMF zero crossing point can be extracted directly from the motor terminal voltage by properly choosing the PWM and sensing strategy. The motor terminal voltage is directly fed into the microcontroller through current-limiting resistors. The resulting feedback signal is not attenuated or filtered. As a result, a sensorless BLDC driver with a much wider speed range from start-up to full speed is obtained [38]. This microcontroller-based sensorless BLDC drive system could be successfully applied to automotive returnless fuel pump applications, in which a BLDC motor life span is typically around 15,000 h, extending the life of the motor almost three-fold. Once a microcontroller is used to perform the brushless commutation, other features can be incorporated into the application, such as electronic returnless, fuel system control, fuel level processing, and fuel tank pressure. These added features simplify the vehicle systems as well as drive overall system cost down.

5.3.5. Application Specific Circuits (ASIC)

Several integrated circuits have been developed to enable sensorless operation of the BLDC. These included Allegro's A8902CLBA [86] or Fairchild Semiconductor's ML4425 [87]. Each of these devices used *back-EMF methods and open-loop starting* [41].

The commercial application specific integrated circuit (ASIC) ML4425 [35] is often used for sensorless control of BLDC drives. It integrates the terminal voltage of the unenergized winding that contains the phase EMF information, and its PLL ensures that the integration result be zero. Thus, the BLDC motor can be commutated with a proper frequency and phase angle. In most cases, the ASIC provides very reliable operation, and its peripheral circuit is simple. Usually, the major problem that the ASIC encounters is the inferior starting performance as open-loop starting procedure is applied.

However, in some applications of BLDC motors, such as driving an air compressor, it requires low starting torque and the open-loop starting can be easily realized. Nevertheless, the commutation of the BLDC drive is significantly retarded during high-speed operation. This is because a relatively wide voltage pulse due to the free-wheeling diode conduction appears in the terminal voltage of the unenergized winding, and it is also integrated by the ASIC.

To overcome the problem, the ASIC should integrate the third harmonic back-EMF instead of the terminal voltage using a voltage integrator and a PLL to process the third harmonic EMF. This solution is more robust to the signal noise [17], and the power loss is reduced at high speed (*i.e.*, 120 krpm) [25].

A practical application of the ML4425 ASIC is to drive electric compressors for automotive HVAC systems. Unlike conventional vehicles that use a belt to drive air conditioner compressors, hybrid electric vehicles use an electric compressor, which is independent from the engine. Because of the requirement of sealing and cost, sensorless drive is embedded in the semi-integrated packaging compressor drive. This ASIC controller is utilized based on the terminal voltage sensing method, providing sequential commutation pulses to the inverter and, hence, *does not need a position sensor* for commutating the motor phases. The commutation is achieved by means of the PLL and speed ranges from 600 to 6,000 rpm can be reached for a 42 V automotive compressor drive [8].

## 6. Applications of BLDC Motor Controllers

As explained briefly in the previous sections, BLDC motors find many applications in every segment of the market, so their control techniques are very important. Automotive, appliance, industrial controls, automation, aviation and so on have applications for BLDC motors. The applications of BLDC motor control can be categorized into three major fields: constant loads, varying loads, and positioning applications [10].

### *6.1. Applications with Constant Loads*

These are the types of applications where a variable speed is more important than keeping the accuracy at a set speed. In addition, the acceleration and deceleration rates are not dynamically changing. In these applications the load is directly coupled to the motor shaft, and this happens in fans [27], pumps, air compressors [1,28] and blowers, which demand low-cost controllers [20], mostly operating in open-loop.

A brushless DC motor built in a compressor of air conditioner is a typical application and it confirms the feasibility and the validity of some sensorless algorithms, such as a variation of the back-EMF zero crossing detection method. It is necessary to modulate the capacity of room air conditioners in proportion to the load results in energy saving and a comfortable environment. The speed of the brushless motor with a permanent magnet rotor can be easily controlled over a wide range by changing the motor voltage. Nevertheless, this type of motor needs a rotor position sensor, and this reduces the system ruggedness and complicates the motor configuration. In particular, the motor speed control and elimination of mechanical sensors are the main points of the sensorless methods, which contribute to the motor built in a completely sealed compressor and make it possible to mass produce. Mechanical sensors have low reliability in high-temperature and the need of a extra hermetic terminal of the sensor signal lead wires, and can be substituted by low-pass filters and voltage comparators [1].

An example of the Terminal Voltage Sensing algorithm can be implemented to drive an IPM BLDC motor which is in a completely sealed compressor of air conditioner. The drive works from about 500 to 7,500 rpm and at low speed, where the amplitude of back-EMF is nearly zero, taking into account that the variation of neutral voltage includes the information of rotor position [28]. Also, the control implementation of a brushless motor, which consists of a three-phase star-connected stator and a four-pole permanent magnet rotor, can be commutated using a 120-electrical-degree type inverter with a capacity of 1.5 kVA, a single-chip microcomputer and comparators [1]. This system has been used for the drive of brushless motor in compressors, and the room air conditioners that contain these control system have been mass-produced since 1982 without any particular problem since then.

*6.2. Applications with Varying Loads*

In these applications the load on the motor varies over a speed range and may demand high-speed control accuracy and good dynamic responses. Home appliances such as washers, dryers and refrigerators are good examples. Also, fuel pump control [38], electronic steering control, engine control and electric vehicle control [8] are examples of these in the automotive industry. In aerospace, there are a number of applications, such as centrifuges, pneumatic devices with electroactuators [3], pumps, robotic arm controls, gyroscope controls and so on. These applications may use speed feedback devices and may run in semi-closed loop or in total closed loop by using advanced control algorithms which complicates the controller and increases the price of the complete system.

The brushless DC motor is well suited for automotive returnless fuel pump applications today, because they are inherently more reliable, more efficient, and with current electronics technology, more cost effective than the standard brush-type fuel-pump motor and controller. In a returnless fuel system, the fuel pump speed is adjusted to maintain constant fuel pressure over the fuel demand/load range. It uses a sensing method that detects the true back-EMF zero crossing point, which requires neither a manufactured neutral voltage nor a great amount of filtering, and provides a wider speed range from startup to full speed. Taking into account that over the last decade, there has been a steady improvement in electronics, control algorithms, and motor technologies, BLDC motors are the preferred solution in not just automotive fuel pumps but also in a broad range of applications using adjustable speed motors, such as home appliances for compressors, air blowers, vacuum cleaners or engine cooling fans. For instance, a brush type fuel pump motor is designed to last 6,000 h, so in certain fleet vehicles this can be expended in less than one year. A BLDC motor life span is typically around 15,000 h, extending the life of the motor by almost three times. Also, because of a microcontroller is used to perform the brushless commutation other features can be incorporated into the application, such as electronic returnless fuel system control, fuel level processing, and fuel tank pressure. These added features simplify the vehicle systems as well as drive overall system cost down [38].

Another important application of BLDC motors is the aerospace field. The implementation of any new technology in the aerospace industry makes strict demands on the safety and reliability of on-board equipment. A fundamental element that drives the EHA/EMA actuator is the DC motor. Nowadays, BLDC motors are widely used, mainly because of their better characteristics and performance. In addition, the ratio of torque delivered to the size of the motor is higher, making it useful in applications where space and weight are critical factors, especially in aerospace. The operation of an actuator is safety critical and thus the actuator requires a reliable control algorithm that ensures safe start-up and operation of the BLDC motor that drives the actuator. Intelligent EHA/EMA actuators and smart actuation systems promising technologies for future power optimised aircraft, adding important benefits in energy consumption, weight savings, easy assembly procedures and maintenance [3].

### *6.3. Positioning Applications*

Most of the industrial and automation types of applications in this category have some kind of power transmission, which could be mechanical gears, gear pump units [4] or timer belts, or a simple belt driven system. The dynamic response of speed and torque are also important, and these applications may also have frequent reversal of rotation direction. A typical cycle will have phases of acceleration, constant speed, and deceleration and positioning. The load on the motor may vary during all of these phases, causing the controller to be complex, such as in applications with hard disk drives (HDDs) [32,79] or DVDs [48,51]. These systems mostly operate in closed loop, and there could be three control loops functioning simultaneously: torque, speed and position. *Optical encoder* or *synchronous resolvers* are used for measuring the actual speed of the motor. In some cases, the same sensors are used to get relative position information. Otherwise, separate position sensors may be used to get absolute positions. Also, computer numeric controlled machines, machines tool [39], industrial processes, and conveyer controls have plenty of applications in this category.

Hydraulic systems are commonly used in automotive applications, since they allow developing higher forces and torques compared with purely electric actuators. For example, passenger cars are equipped with hydraulically-assisted brakes, clutches and power steering systems, while in commercial vehicles hydraulic power is used to operate also lifting systems and other auxiliary machineries. If a variable speed electric motor is coupled to the hydraulic pump, a flow control valve is always necessary, and the possibility to regulate the rotational speed of the pump *i*ndependently from the engine speed allows a significant reduction of parasitic losses. However, this solution requires the design of a specific electronic controller, capable of motor speed regulation according to the hydraulic load dynamic requirements. Moreover, since the power supply of these electro-hydraulic systems must be the battery of the vehicle, to allow operation even when the engine is switched off, electric motors and power electronics must be designed for low-voltage and high-current ratings, especially in commercial vehicle applications [4]. Then, a brushless motor are the appropriate device to this purpose.

On the one hand, taking into account these considerations, the design of the electronic part of a Motor Pump Unit (MPU) is composed by a hydraulic gear pump, a permanent magnet brushless motor (chosen because of its advantages over brushed DC motors) and a power converter with a microcontroller unit, which implements the sensorless speed control scheme. For example, the key features of a MPU for commercial vehicle applications can be the following: 24 V supply, 2 kW maximum output power, 4,000 rpm maximum motor speed, 10 Nm maximum motor torque, and 150 bar maximum hydraulic pressure [4]. On the other hand, the main issues in the back-EMF zero-crossing detection method applied in the system are related to noise superimposed on both phase voltage and mid-point voltage, because of PWM modulation of the power converter and the particular care for the implementation of start-up algorithms. A more efficient solution can be implemented if a digital controller was able to sample and digitize back-EMF measurements synchronously with PWM modulation (*i.e.*, during PWM off periods) and implement more robust zero-crossing detection algorithms [88]. Moreover, phase advance strategies will allow to extend the speed range of the motor, which could also be more easily implemented on a high-performance motor control signal processor, such as a Microchip dsPIC30F6010 Digital Signal Controller using their PWM generators and analog-to-digital converters [89].

Another important application in this category is the hard disk drives or HDDs. HDDs tend to have high spin speeds in order to reduce the access time in data reading and writing. The highest spin speed of commercial HDDs has reached 15,000 rpm and will be higher in the near future. However, for the small form factor HDDs, the back-EMF amplitudes of their spindle motors are becoming low even at the rated speed, and some methods based on the zero-crossing detection of back EMFs does not work well when the terminal voltage spikes last relatively longer at higher spin speed or the phase back- EMF amplitude is very small [32]. Recently, due to HDDs being widely used in mobile applications, the power-supply voltage has been reduced and the detection of the rotor position from the back EMF is difficult at low speed. Thus, the stable starting and acceleration to nominal operating speed, regardless of severe mechanical disturbance is the utmost concern in these applications [79]. However, since the 1970s, many methods have been developed for solving the problems in the sensorless rotor position detection, such as the back-EMF integration method or digital filtering procedures, which have been developed to identify the true and false ZCPs of phase back EMFs caused by the terminal voltage spikes due to the residual phase currents during commutations. Also, for small power applications of BLDC motor drives, such as digital video disks, due to the use of battery or/and limited space for heat dissipation, reduction of power consumption becomes one of the main concerns for the development of PWM techniques, which have been developed for controlling power devices by means of variable voltage and frequency [51].

## 7. Conclusions

In this paper a review of position control methods for BLDC motors has been presented. The fundamentals of various techniques have been introduced, mainly back-EMF schemes and estimators, as a useful reference for preliminary investigation of conventional methods. Advances in the position control and applications were also discussed.

To provide insight in control techniques and their benefits a classification of existing methods and newer methods were presented with their merits and drawbacks. From the above discussion, it is obvious that the control for BLDC motors using position sensors, such as shaft encoders, resolvers or Hall-effect probes, can be improved by means of the elimination of these sensors to further reduce cost and increase reliability. Furthermore, sensorless control is the only choice for some applications where those sensors cannot function reliably due to harsh environmental conditions and a higher performance is required.

## Acknowledgements

This work was supported by the regional 2010 Research Project Plan of Junta de Castilla y León (Spain), reference project VA034A10-2. This work was also possible thank to the collaboration of Ernesto Vazquez-Sanchez through the Contratación de Personal Investigador de Reciente Titulación program, which was financed by Consejería de Educación of Junta de Castilla y León (Spain) and cofinanced by European Social Fund.